\documentclass[11pt,reqno]{amsart}
\usepackage{amsfonts}
\usepackage{mathrsfs}
\usepackage{amsmath}
\usepackage{amssymb,amsfonts}

\newtheorem{thm}{Theorem}[section]
\newtheorem{lem}{Lemma}[section]
\newtheorem{defn}{Definition}[section]

\numberwithin{equation}{section}
\newtheorem{rmk}{Remark}[section]

\def\pf{{\textit {Proof:} }}

\newcommand{\mysection}[1]{\section{#1}\setcounter{equation}{0}}

\newfont{\bb}{msbm10 at 11pt}

\newcommand{\bal}{\begin{aligned}}      \newcommand{\eal}{\end{aligned}}
\newcommand{\ba}{\begin{array}}      \newcommand{\ea}{\end{array}}
\newcommand{\bc}{\begin{center}}     \newcommand{\ec}{\end{center}}
\newcommand{\be}{\begin{enumerate}}  \newcommand{\ee}{\end{enumerate}}
\newcommand{\beq}{\begin{eqnarray}}  \newcommand{\eeq}{\end{eqnarray}}
\newcommand{\beQ}{\begin{eqnarray*}} \newcommand{\eeQ}{\end{eqnarray*}}
\newcommand{\bi}{\begin{itemize}}    \newcommand{\ei}{\end{itemize}}
\newcommand{\bt}{\begin{tabular}}    \newcommand{\et}{\end{tabular}}
\newcommand{\bdm}{\begin{displaymath}} \newcommand{\edm}{\end{displaymath}}
\newcommand{\pt}{\partial}
\newcommand{\pu}{\partial u}
\newcommand{\pr}{\partial r}
\newcommand{\rw}{\rightarrow}

\newcommand{\lrw}{\longrightarrow}
\newcommand{\Lrw}{\Longrightarrow}

\newcommand{\bhi}{\bar{h}_{n_i}}
\newcommand{\pv}{\partial V}
\newcommand{\pb}{\partial\beta}

\newcommand{\ep}{\epsilon}
\newcommand{\ir}{\int_{0}^{r}}

\newcommand{\Ba}{\Big |}
\newcommand{\ani}{A_{n_i}}

\newcommand{\hni}{h_{n_i}}

\newcommand{\uz}{u_0}
\newcommand{\cni}{\chi_{n_i}}

\newcommand{\ift}{\infty}

\newcommand{\ha}{h_{\alpha}}
\newcommand{\bha}{\bar{h}_{\alpha}}
\newcommand{\ga}{g_{\alpha}}
\newcommand{\bga}{\bar{g}_{\alpha}}
\newcommand{\Bl}{\Big (}
\newcommand{\al}{\alpha}
\newcommand{\bhn}{\bar{h}_n}
\newcommand{\bgn}{\bar{g}_n}
\newcommand{\pph}{\partial \phi}
\newcommand{\Br}{\Big )}
\newcommand{\ur}{\Big(1+\frac{tu}{2}+r\Big)}
\newcommand{\uu}{\Big(1+\frac{tu}{2}\Big)}
\newcommand{\gni}{g_{n_i}}
\newcommand{\bgi}{\bar{g}_{n_i}}
\newcommand{\phn}{\partial h_n}
\newcommand{\bg}{\bar{g}}
\newcommand{\Blb}{\Big\{ }
\newcommand{\Brb}{\Big\}}
\newcommand{\bh}{\bar{h}}
\newcommand{\hai}{h_{\alpha_i}}
\newcommand{\cn}{\chi_n}

\newcommand{\ma}{M_{0,\alpha}}

\newcommand{\ph}{\partial h}
\newcommand{\ppp}{\partial^2\phi}
\newcommand{\prr}{\partial r^2}

\newcommand{\dg}{\dot{\gamma}}

\def\qed{\hfill{Q.E.D.}\smallskip}

\newcommand{\ls}{\setlength{\baselineskip}{12pt}
                 \setlength{\parskip}{3mm}}

\begin{document}

\title[Einstein-scalar-field equations]{Spherically symmetric Einstein-scalar-field equations for wave-like decaying null infinity}
\let\thefootnote\relax\footnotetext{This work is partially the first author's PhD thesis at the University of Chinese Academy of Sciences.}
\author[C Liu]{Chuxiao Liu$^{\dag}$}
\address[]{$^{\dag}$School of Mathematics and Information Science, Guangxi University, Nanning, Guangxi 530004, PR China}
\email{cxliu@gxu.edu.cn}
\author[X Zhang]{Xiao Zhang$^{\flat}$}
\address[]{$^{\flat}$Guangxi Center for Mathematical Research, Guangxi University, Nanning, Guangxi 530004, PR China}
\address[]{$^{\flat}$Academy of Mathematics and Systems Science, Chinese Academy of Sciences, Beijing 100190, PR China}
\address[]{$^{\flat}$School of Mathematical Sciences, University of Chinese Academy of Sciences, Beijing 100049, PR China}
\email{xzhang@gxu.edu.cn, xzhang@amss.ac.cn}

\subjclass[2000]{53C50, 58J45, 83C05}
\keywords{Einstein-scalar-field equations; Spherically symmetric Bondi-Sachs metrics; null infinity; wake-like decaying conditions}

\date{}

\begin{abstract}
We show that the spherically symmetric Einstein-scalar-field equations for wave-like decaying initial data at null infinity have unique
local solutions and unique global solutions for small initial data. We also generalize Christodoulou's global generalized solutions to the
wave-like decaying initial data. We emphasize that this decaying condition is sharp.
\end{abstract}

\maketitle \pagenumbering{arabic}

\mysection{Introduction}\ls

\subsection{Spherically symmetric Einstein-scalar-field equations}\ls

Let $g$ be a 4-dimensional Lorentzian metric, $\phi$ be a real function. The Einstein-scalar-field equations are the following systems
\beq\label{e-s-f}
\begin{matrix}
\left\{
\begin{aligned}
& R_{\mu\nu}-\frac{1}{2} R g_{\mu \nu} =8\pi T_{\mu \nu}\\
& T_{\mu \nu}=\partial_{\mu} \phi \partial_{\nu} \phi -\frac{1}{2} g_{\mu \nu} \partial ^\alpha \phi \partial _\alpha \phi
\end{aligned}
\right.
\end{matrix}
\Longleftrightarrow R_{\mu\nu}=8\pi \partial_{\mu} \phi \partial_{\nu} \phi.
\eeq

The spherically symmetric solutions of (\ref{e-s-f}) were studied extensively by Christodoulou \cite{C1, C2, C3, C4, C5, C6}
using slight different notation of the following spherically symmetric Bondi-Sachs metrics \cite{BBM, S}
\beq
g=-e^{2\beta(u,r)}\frac{V(u,r)}{r}du^2-2e^{2\beta (u,r)}dudr+r^2 \big(d\theta^2+\sin^2\theta d\psi^2 \big), \label{1}
\eeq
where $u$ is referred as the retarded time, $$V(u,r)>0,$$ and $\beta$, $V$ satisfy the following boundary conditions
\beq
\beta (u, \infty) =0, \quad \lim_{r \rw \infty} \frac{V(u,r)}{r}=1. \label{b-c}
\eeq
Using the expressions of the Ricci curvature (c.f. Appendix), we obtain the Einstein-scalar-field equations
\beq
\begin{matrix}
\left\{
\begin{aligned}
-&\frac{V}{r}\frac{\partial\beta}{\partial u}+\frac{1}{2r}\frac{\partial V}{\pt u}
                =8\pi \Big[\Bl\frac{\pt \phi}{\pu}\Br^2-\frac{V}{r}\frac{\pt \phi}{\pu}\frac{\pt \phi}{\pr}\Big],\\
-&2\frac{\partial^2\beta}{\partial u\partial r}+\frac{V}{r}\frac{\partial^2\beta}{\partial r^2}
                +\frac{1}{r}\frac{\partial \beta}{\partial r}\frac{\partial V}{\partial r}+\frac{V}{r^2}\frac{\partial \beta}{\partial r}
                =8\pi \frac{\pt \phi}{\pu}\frac{\pt \phi}{\pr},\\
 &\frac{\pb}{\pr}=2\pi r \Bl\frac{\pt\phi}{\pr}\Br^2,\quad \frac{\pv}{\pr}=e^{2\beta}.
\end{aligned}\label{2-0}
\right.
\end{matrix}
\eeq
The function $\phi$ satisfies the wave equation by the twice contracted Bianchi identity
\beq
-2\Bl\frac{\ppp}{\pu\pr}+\frac{1}{r}\frac{\pph}{\pu}\Br+\frac{V}{r}\frac{\ppp}{\prr}+\Bl\frac{V}{r^2}
+\frac{1}{r}\frac{\pv}{\pr}\Br\frac{\pph}{\pr}=0.\label{2}
\eeq

Given function $f(u,r)$, denote $\bar f (u,r)$ its average
\beQ
\bar{f}(u,r)=\dfrac{1}{r}\int_{0}^{r}{f(u,r')dr'}.
\eeQ
Denote $D$ the derivative along the incoming light rays parameterized by $u$
\beQ
D=\dfrac{\pt}{\pu}-\dfrac{V}{2r}\dfrac{\pt}{\pr}.
\eeQ
In \cite{C1}, Christodoulou introduced
\beQ
h=\frac{\partial (r \phi)}{\partial r}
\eeQ
and showed that the spherically symmetric Einstein-scalar-field equations (\ref{2-0}), (\ref{2}) under boundary
conditions (\ref{b-c}) are equivalent to the system \cite{C1}
\beq
\begin{matrix}
\left\{
\begin{aligned}
\beta &=-2\pi\int_{r}^{\ift}{(h-\bar{h})^2\frac{dr'}{r'}},\\
g &=e^{2\beta}, \quad \bar{g}=\frac{1}{r}\ir{g dr'}=\frac{V}{r},\\
D h&=\frac{1}{2r}(g-\bar{g})(h-\bar{h}).   \label{3}
\end{aligned}\right.
\end{matrix}
\eeq

Let $r=\chi(u)$ be the integral curves of $D$. They are the incoming light rays called characteristics and satisfy the ordinary differential equation \cite{C1}
\beq
\frac{dr}{du}=\frac{d\chi(u)}{du}=-\frac{1}{2}\bg(u,r). \label{light-ray}
\eeq
%
%
\subsection{The Bondi mass and the Bondi-Christodoulou mass}\ls

Denote
\beq
m_B(u) =\frac{r}{2}\Big(1-\frac{V}{r} \Big)=\frac{r}{2}(1-\bar{g}). \label{bondi}
\eeq
Following from Bondi \cite{BBM}, the Bondi mass and the final Bondi mass are defined as
\beQ
M_B(u) =\lim_{r\rw\ift} m_B(u), \quad M_{B1}=\lim _{u \rightarrow \infty} M_B(u).
\eeQ

In \cite{C1}, Christodoulou defined a mass function
\beq
m(u,r)=\frac{r}{2}\Bl1-\frac{\bar{g}}{g}\Br   \label{C-H}
\eeq
and proved that it satisfies
\beq
\begin{matrix}
\left\{
\begin{aligned}
\frac{\pt m}{\pr}&=2\pi \frac{\bg}{g}(h-\bh)^2 \geq 0,\\
D m(u,r)&=-\frac{\pi}{g} \Big(\int_0 ^r \bar{g} (h-\bar{h})^2  \frac{d r'}{r'}\Big)^2.
\end{aligned}  \label{m}
\right.
\end{matrix}
\eeq
He also defined
\beQ
M(u)=\lim _{r \rightarrow \infty} m(u,r),\quad M _1=\lim _{u \rightarrow \infty} M(u).
\eeQ
In general, $M(u)$ may not be equal to $M_B(u)$. So it is reasonable to call $M(u)$ the Bondi-Christodoulou mass and $M_1$ the final Bondi-Christodoulou mass.

If the metric $g$ is regular in $[0, \infty)$ at each $u$, then (\ref{3}), (\ref{m}) give
\beQ
g\leq 1, \quad m(u,r)\geq m(u,0)=0.
\eeQ
It follows that
\beQ
m_B(u) \geq m(u), \quad M_B(u) \geq M(u), \quad M_{B1} \geq M_1.
\eeQ

\subsection{Existence and uniqueness}\ls

Throughout the paper, we always assume
\beQ
0< \ep \leq 2.
\eeQ
We summarize the main results proved by Christodoulou in \cite{C1, C2, C3, C4, C5} as the following three theorems.
(They are proved for $\ep = 2$ in the papers. But the proofs can be extended to the case $0< \ep \leq 2$, see \cite{LO}.)

\begin{thm}
For any initial data $\breve{h}(r)\in C^1[0,\ift)$ which satisfies
\beQ
\breve{h}(r)=O\Big(\frac{1}{r^{1+\ep} }\Big), \quad \frac{\partial \breve{h}}{\partial r}(r) =O\Big(\frac{1}{r^{2+\ep} }\Big)
\eeQ
as $r \rightarrow \infty$, there exists $\uz>0$ and a unique classical solution
\beQ
h(u,r) \in C^1 \Big([0,\uz]\times [0,\ift)\Big)
\eeQ
of (\ref{3}) which satisfies the initial condition $h(0,r)=\breve{h}(r)$ and
\beQ
h(u, r)=O\Big(\frac{1}{r^{1+\ep} }\Big), \quad \frac{\partial h}{\partial r}(u,r) =O\Big(\frac{1}{r^{2+\ep} }\Big)
\eeQ
uniformly in $u\in [0, u_0]$ as $r \rightarrow \infty$.
\end{thm}

\begin{thm}
For any initial data $\breve{h}(r)\in C^1[0,\ift)$ which satisfies
\beQ
\breve{h}(r)=O\Big(\frac{1}{r^{1+\ep} }\Big), \quad \frac{\partial \breve{h}}{\partial r}(r) =O\Big(\frac{1}{r^{2+\ep} }\Big)
\eeQ
as $r \rightarrow \infty$, there exists $\delta _0>0$ such that if
\beQ
\inf _{b>0} \sup _{r\geq 0} \Big\{\Big(1+\frac{r}{b}\Big)^{1+\ep} |h(r)| + \Big(1+\frac{r}{b}\Big)^{2+\ep} \Big|b \frac{\partial h}{\partial r}(r)\Big|   \Big\}\leq \delta _0,
\eeQ
there exists a global classical solution
\beQ
h(u,r) \in C^1 \Big([0,\infty)\times [0,\ift)\Big)
\eeQ
of (\ref{3}) which satisfies the initial condition $h(0,r)=\breve{h}(r)$ and
\beQ
\big| h(u,r) \big| \leq \frac{C}{(1+u+r)^{1+\ep}}, \quad \Big| \frac{\partial h}{\partial r}(u,r)\Big| \leq \frac{C}{(1+u+r)^{2+\ep}}.
\eeQ
The corresponding spacetime is future causally geodesically complete with vanishing final Bondi-Christodoulou mass $M_1$.
Furthermore, if the final Bondi-Christodoulou mass $M_1$ is positive, a black hole forms and, in the region exterior to the Schwarzschild sphere $r=2M_1$, the spacetime metric tends to the Schwarzschild metric.
\end{thm}

In \cite{C2, C3}, Christodoulou introduced the following definition of global generalized solutions and proved their existence and uniqueness.

\begin{defn}\label{gensolu}
Denote $I=[0,\infty)\times(0,\ift)$ the complement of the central line. A generalized solution to the equation (\ref{3}) is a function
\beQ
h(u,r) \in C^1(I)
\eeQ
such that, for each $u \in [0, \infty)$, $h(u,r) \in L^2(0,\ift)$ and the quantity $\int_{0}^{\ift}{h^2dr}$ is bounded by a continuous function of $u$. Moreover, $h$ satisfies the following properties.
\bi
\item[(1)] $h$ satisfies the equation (\ref{3}) in $I$ and takes $h(0,r)=\breve{h}$ at $u=0$ as the initial data, $\bh$, $g$, $\bg$ are continuous functions in $I$.
\item[(2)] At each $u$, for arbitrary $r_1$, $$\dfrac{g}{\bg}\in L^1(0,r_1).$$
\item[(3)] For almost all $u$, and any $(u_1,r_1)\in I$,
\beQ
\xi=\lim\limits_{\delta\rw 0}{\int_{\delta}^{r}{\bg(h-\bh)\frac{dr'}{r'}}}
\eeQ
exists and $$\dfrac{g^{\frac{1}{2}}\xi}{\bg r^{\frac{1}{2}}}\in L^2\Big((0,u_1)\times(0,r_1)\Big).$$
\item[(4)] $\bh$, $m$ are weakly differentiable in $I$, and
\beQ
D\bh=\dfrac{\xi}{2r}, \quad Dm=-\dfrac{\pi }{g}\xi^2.
\eeQ
\item[(5)] For each $(u_1,r_1)\in I$, denote $r_0=\chi_{u_1}(0;r_1)$,
$$I(u_1,r_1)=\Big\{(u,r)|0<r<\chi_{u_1}(u;r),\,\,0<u<u_1\Big\}.$$
Then the following integral identity holds
$$
\int_{0}^{r_1}{\frac{g}{\bg}(u_1,r)dr}+2\pi \iint\limits_{I(u_1,r_1)}{\frac{g\xi^2}{\bg^2r}drdu}+\frac{1}{2}\int_{0}^{u_1}g(u,0)du=\int_{0}^{r_0}{\frac{g}{\bg}(0,r)dr}.
$$
\ei
\end{defn}

\begin{thm}
For any initial data $\breve{h}(r)\in C^1[0,\ift)$ which satisfies
\beQ
\breve{h}(r)=O\Big(\frac{1}{r^{1+\ep} }\Big), \quad \frac{\partial \breve{h}}{\partial r}(r) =O\Big(\frac{1}{r^{2+\ep} }\Big)
\eeQ
as $r \rightarrow \infty$, there exists at least one global generalized solution which has the same data as a classical solution coincides with it in the domain of existence of the latter.
\end{thm}

On the other hand, by using the double null coordinate system
\beq
g=-\Omega ^2 du dv +r^2 \big(d\theta^2+\sin^2\theta d\phi^2 \big), \label{1-1}
\eeq
Christodoulou solved the characteristic (lightlike) initial value problem for small bounded variation norm and proved the following theorem \cite{C6}.
\begin{thm}
If there exists universal constant $\delta _1 >0$ such that
\beQ
\int _{u_0} ^\infty \Big| \frac{\partial h}{\partial v} (u_0, v) \Big| dv  \leq \delta _1,
\eeQ
then the unique spherically symmetric global solution exists for metric (\ref{1-1}).
\end{thm}

This result was extended to more general cases by Luk-Oh and Luk-Oh-Yang \cite{LO, LOY}. In particular, Luk, Oh and Yang proved the following theorem in \cite{LOY}.
\begin{thm}
The unique symmetrically symmetric global solution exists for metric (\ref{1-1}) if
\beQ
\begin{aligned}
\int _u ^v \big| h(u_0, v') \big| dv'  \leq \epsilon \big(v-u \big)^{1-\gamma}, \quad
|h(u_0, v)| + \Big| \frac{\partial h} {\partial v} (u_0, v)\Big| \leq \epsilon,
\end{aligned}
\eeQ
for any $v \geq u \geq u_0$, where $\gamma >0$ is certain positive constant. Moreover, the resulting spacetime is future causally geodesically complete.
\end{thm}

We remark that the $u$ and $r$-slices in metric (\ref{1}) provide null and timelike hypersurfaces respectively. It yields characteristic (timelike) initial value problem. Thus the issue to solve the Einstein-scalar-field equations in metric (\ref{1}) is different from that in metric (\ref{1-1}), and another characteristic in (\ref{1}) may not be given by level set of certain function $v$.

\subsection{Wave-like decaying null infinity}\ls

Gravitational waves can be described by the Bondi-Sachs metrics
\beQ
-\Big(e^{2\beta}\frac{V}{r}-r^2h_{AB}U^AU^B\Big)du^2-2e^{2\beta}du dr-2r^2h_{AB} U^Bdudx^A+r^2 g_2
\eeQ
where
\beQ
g_2=h_{AB}(u, x^A, x^B) dx^Adx^B
\eeQ
is certain Riemannian metrics on unit 2-sphere. Intuitively, when the waves arrive at infinity, the metrics should decay to the Minkowski metric wave-likely. For instance, as $r \rightarrow \infty$
\beQ
g_{Bondi-Sachs}=g_{Minkowski}+O\Big(\frac{\sin (r)}{(1+r)^{1+\delta }}\Big),\quad \delta >-1.
\eeQ
Note that this kind of error terms satisfy
\beQ
(1+r)^{1+\delta }\Big|\Big(\frac{\partial}{\partial r}\Big)^k \Big(\frac{\sin (r)}{(1+r)^{1+\delta}} \Big) \Big|<C
\eeQ
for any nonnegative integer $k\geq 0$. Motivated by this, the second author posed the wave-like decaying spatial infinity and found that, for the Bondi-Sachs metrics, the past limit of the Bondi energy-momentum are not equal to the ADM total energy-momentum \cite{Z, HZ}
\beQ
\lim _{u \rightarrow -\infty} M_{Bondi,0} \neq E_{ADM}, \qquad  \lim _{u \rightarrow -\infty} M_{Bondi, k} \neq P_{ADM,k},
\eeQ
where $1 \leq k \leq 3$. This violates physicist's expectation that they should be equal to each other, see, e.g. \cite{PR}.

Therefore it is important to investigate wave-like boundary conditions and study new physical properties for gravitational waves. In particular,
for the spherically symmetric Einstein-scalar-field equations, it is interesting to study existence and uniqueness of (\ref{3}) for wave-like decaying initial condition, e.g. $$\breve{h}(r)=\frac{\sin (r)}{(1+r)^{1+\ep }}.$$

\subsection{Main results}\ls

Let $h(u,r)$ be a continuous function and possess a continuous partial derivative with respect to $r$ defined on $[0,u_1]\times [0,\ift)$ if $u_1<\infty$, or, on $[0,\infty)\times [0,\ift)$ if $u_1=\infty$. We call that $h(u,r)$ satisfies the wave-like decaying condition at a retarded time $u$ if
\beq
\sup _{r \geq 0} \Big\{(1+r)^{1+\ep} \Big(|h(u,r)| + \Big|\frac{\partial h}{\partial r} (u,r) \Big| \Big) \Big\}<\infty.\label{wave-like}
\eeq
Here $h(u,r)$ is replaced by $\breve{h}(r)$ for initial data.

In this paper we first prove the following local existence and uniqueness of classical solutions.

\begin{thm}\label{thm1}
For any initial data $\breve{h}(r) \in C^1[0,\ift)$ which satisfies (\ref{wave-like}), then there exists $\uz>0$ and a unique classical solution
\beQ
h(u,r) \in C^1 \Big([0,\uz]\times [0,\ift)\Big)
\eeQ
of (\ref{3}) which satisfies the initial condition $h(0,r)=\breve{h}(r)$ and the decay property (\ref{wave-like}) uniformly in $u\in [0, u_0]$.
\end{thm}

We then prove the following global existence and uniqueness for small initial data.

\begin{thm}\label{thm2}
Consider initial data $\breve{h}(r)\in C^1[0,\ift)$ which satisfies (\ref{wave-like}). Denote
\beQ
d_0 : =\inf_{b>0}\sup_{r\geq0}{\Blb \Big(1+\frac{r}{b}\Big)^{1+\ep} \Bl|h_0(r)|+\Ba b\frac{\pt h_0}{\pr}(r)\Ba\Br\Brb}.
\eeQ
Then there exists $\delta >0$ such that if $d_0 <\delta$, there exists a unique global classical solution
\beQ
h(u,r) \in C^1 \Big([0,\infty)\times [0,\ift)\Big)
\eeQ
of (\ref{3}) which satisfies the initial condition $h(0,r)=\breve{h}(r)$ and the decay property
\beQ
|h(u,r)|\leq \frac{C}{\Big(1+\frac{u}{2}+r\Big)^{1+\ep}},\quad \Ba\frac{\pt h}{\pr}(u,r)\Ba\leq\frac{C}{\Big(1+\frac{u}{2}+r\Big)^{1+\ep}}
\eeQ
for some constant $C$ depending on $\ep$ only. Moreover, the corresponding spacetime is future causally geodesically complete with vanishing final Bondi mass $M_{B1}$.
\end{thm}

Finally, we extend Christodoulou's global generalized solutions to the case of wave-like decaying initial data.

\begin{thm}\label{thm3}
Given initial data $\breve{h}(r) \in C^1[0,\ift)$ which satisfies (\ref{wave-like}), there exists at least one global generalized solution which has the same data as a classical solution coincides with it in the domain of existence of the latter.
\end{thm}

We remark that the wave-like decaying condition is sharp for the well-posedness of Einstein-scalar-field equations and we prove our main theorems by adopting Christodoulou's idea. Due to the slowly wave-like decaying condition, we need to derive some key estimates in more delicate way.

The paper is organized as follows. In Section 2, we prove the three lemmas which contain the main estimates for the solutions of the Einstein-scalar-field equations. In Section 3, we prove Theorem \ref{thm1}. In Section 4, we prove Theorem \ref{thm2}. In Section 5, we prove Theorem \ref{thm3}.

\mysection{Main estimates}\ls

In this section, we follow from Christodoulou's argument in \cite{C1} to prove three lemmas for the wave-like decaying solutions. As the solutions have a slowly decaying partial derivative with respect to $r$, we need to derive much more fine estimates.

Let $t \in [0, 1]$, $u_1 \in [0, \infty)$ or $u_1=\infty$. Denote $X_{u_1,t}$, $Y_{u_1,t}$ the spaces of continuous functions which possess continuous partial derivatives with respect to $r$ defined on $[0,u_1]\times [0,\ift)$ if $u_1<\infty$, or, on $[0,\infty)\times [0,\ift)$ if $u_1=\infty$ such that the following norms are finite
$$\|f\|_{X_{u_1,t}}=\sup_{u\in[0,u_1] \,\text{or}\,\, [0, \infty)}\sup_{r\geq 0}\Big\{\Big(1+\frac{tu}{2}+r \Big)^{1+\ep}
 \Big(|f(u,r)|+\Big|\frac{\pt f}{\pr}(u,r)\Big|\Big)\Big\},$$
$$\|f\|_{Y_{u_1,t}}=\sup_{u\in[0,u_1] \,\text{or}\,\,[0, \infty)}\sup_{r\geq 0}\Big\{\Big(1+\frac{tu}{2}+r \Big)^{1+\ep} |f(u,r)| \Big\}.$$
It is obvious that $X_{u_1,t}$, $Y_{u_1,t}$ are Banach spaces equipped with the corresponding norms.

For any initial data $h(0,r) =\breve{h}(r) \in X_{0,t}$, denote
\beq
\|h\|_{X_{0,t}}=\sup_{r\geq 0}\Big\{(1+r)^{1+\ep}\Bl|h(0,r)|+\Big|\frac{\ph}{\pr}(0,r)\Big|\Br\Big\}=d>0.\label{h0}
\eeq
We construct a sequence of approximate solutions by setting
\beq
h_0(u,r)=h\Big(0,\frac{t u}{2}+r\Big). \label{h_0}
\eeq
For $h_n\in X_{u_1,t}$, define $h_{n+1}$ to be the solution of the equation
\beq
D_n h_{n+1}-\frac{1}{2r}(g_n-\bg_n)h_{n+1}=-\frac{1}{2r}(g_n-\bg_n)\bh_n \label{10}
\eeq
with the initial condition
\beQ
h_{n+1}(0,r)=h(0,r),
\eeQ
where $g_n$ is the $g$-function corresponding to $h_n$ and $D_n$ is the $D$-operator corresponding to $g_n$.

Denote
\beq
&&\quad \,\,\,\,\,c=\left\{\begin{aligned}
      &\frac{6}{\ep|1-\ep|},&\qquad  \qquad \qquad \ep\neq 1,\\
      &24,& \qquad \qquad \qquad \ep=1,
\end{aligned}
\right. \label{A5}\\
&&\quad \,\,\, c_t=\left\{\begin{aligned}
&1,\qquad \qquad \qquad \qquad \quad \,\,\,\, t=0,\\
&\exp\Big(\frac{2(1+\ep)\pi c^2x^2}{3}\Big),\quad 0<t\leq 1,\\
\end{aligned}
\right.\label{A9}\\
&&\,\, a(u)=\left\{\begin{aligned}
&u,\,\,\qquad \qquad \qquad \qquad \quad \,\,u<\infty,\\
&\frac{1}{t\ep},\qquad \qquad \qquad \qquad \quad \,u=\infty, \,\ t \neq 0,
\end{aligned}
\right.\label{A10}
\eeq
and
$$F_{t}(u,x)=c_t^2\Bl2 +8\pi c^2x^2 a(u)\Br\Bl d+\frac{24+6\ep+4\pi}{3\ep}c^3x^3 a(u)
\Br\exp{\Bl\frac{8\pi c^2 x^2a(u)}{3}\Br},$$
$$C_{t}(u,x) =\frac{32\pi (\ep +1)}{\ep ^{2}} c_t c^3 x^2 a(u) (3+4\pi c^2x^2) \exp\Big({\frac{4\pi c^2x^2a(u)}{3}}\Big).$$

\begin{lem}\label{lemA}
Let initial data $\breve{h}(r)$ satisfy (\ref{h0}). If there exists some constant $x>0$ such that
\beQ
\|h_n(u,r)\|_{X_{u_1,t}}\leq x,
\eeQ
then it satisfies that
\beq
\|h_{n+1}(u,r)\|_{X_{u_1,t}}\leq F_{t}(u_1,x). \label{A20}
\eeq
\end{lem}
\pf First we prove the following claim case by case.\\\\
{\em Claim I}: For $0<\ep\leq 2$,
\beq
|(h_n-\bhn)(u,r)|\leq\frac{c x r\uu^{1-\ep}}{\ur^2},\label{A4}
\eeq
where $c$ is given by (\ref{A5}).

Indeed, it is easy to find
\beq
\begin{aligned}
|\bhn(u,r)|&\leq \frac{1}{r}\ir{|h_n(u,r)|dr}\\
&\leq \frac{x}{\ep r}\Big[\frac{1}{\Big(1+\frac{tu}{2}\Big)^{\ep}}-\frac{1}{\Big(1+\frac{tu}{2}+r\Big)^{\ep}}\Big]\\
&\leq \frac{x}{\ep r \Big(1+\frac{tu}{2}\Big)^{\ep}}  \Big[1-\frac{\Big(1+\frac{tu}{2}\Big)^2}{\Big(1+\frac{tu}{2}+r\Big)^2} \Big]\\
&\leq \frac{2x}{\ep\Big(1+\frac{tu}{2}\Big)^{\ep}\Big(1+\frac{tu}{2}+r\Big)}.
\label{A1}
\end{aligned}
\eeq\\
{\it Case I}$\,$: $0<\ep<1$.

(i) For $r\geq 1+\dfrac{tu}{2}$, we have
\beq
\begin{aligned}
|(h_n-\bhn)(u,r)|&\leq |h_n|+|\bhn|\\
&\leq \frac{x}{\Big(1+\frac{tu}{2}+r\Big)^{1+\ep}}+\frac{2x}{\ep \uu^{\ep}\ur}\\
&\leq\frac{6x r}{\ep \uu^{\ep}\ur^2}\\
&\leq\frac{6x r\uu^{1-\ep}}{\ep\ur^2}.
\label{A2}
\end{aligned}
\eeq

(ii) For $0\leq r\leq 1+\dfrac{tu}{2}$, we have
\beQ
\begin{aligned}
|(h_n&-\bhn)(u,r)|\leq \frac{1}{r}\ir{\int_{r'}^{r}{\Ba\frac{\pt h_n}{\pt s}\Ba ds}dr'}\\
&\leq \frac{x}{\ep r}\ir{\Big[\frac{1}{\Big(1+\frac{tu}{2}+r'\Big)^{\ep}}-\frac{1}{\ur^{\ep}}\Big]dr'}\\
&\leq \frac{x}{\ep(1-\ep)r}\Big[\ur^{1-\ep}-\uu^{1-\ep}-\frac{(1-\ep)r}{\ur^{\ep}}\Big]\\
&\leq\frac{x\uu^{1-\ep}}{\ep(1-\ep)r \ur^2}\Big[F(r)+5r^2\Big],
\end{aligned}
\eeQ
where
\beQ
\begin{aligned}
F(r)=&\ur^{3-\ep}\uu^{\ep-1}-\ur^2\\
     &-(1-\ep)r\uu-5r^2.
\end{aligned}
\eeQ

By analyzing the monotonicity of $F(r)$, we obtain
\beQ
\begin{aligned}
F^{'''}\geq 0 & \Longrightarrow F^{''}(r)\leq F^{''}\Big(1+\frac{tu}{2}\Big)=(3-\ep)(2-\ep)2^{1-\ep}-12 \leq 0 \\
              & \Longrightarrow F^{'}(r)\leq F^{'}(0)=0 \Longrightarrow F(r)\leq F(0)=0.
\end{aligned}
\eeQ
This gives
\beq
|(h_n-\bhn)(u,r)|\leq\frac{5x r\uu^{1-\ep}}{\ep(1-\ep)\ur^2}.
\label{A3}
\eeq
Thus, (\ref{A2}) and (\ref{A3}) give that, for any $r\geq 0$,
\beQ
|(h_n-\bhn)(u,r)|\leq\frac{6x r\uu^{1-\ep}}{\ep(1-\ep)\ur^2}.
\eeQ\\
{\it Case II}$\,$: $\ep=1$.

Since
\beQ
\Big|\frac{\pt h_n}{\pt r}\Big|\leq\frac{x}{\ur^2}\leq \frac{x}{\ur^{1+\frac{1}{2}}}.
\eeQ
Then we can use the estimate for $\ep=\frac{1}{2}$ in ${\it Case\,I}$ and obtain
\beQ
|(h_n-\bhn)(u,r)|\leq\frac{24x r\uu^{1-\ep}}{\ur^2}.
\eeQ\\
{\it Case III}$\,$: $1<\ep\leq 2$.

Using the same argument as in {\it Case I} and choose $F$ as follows.
\beQ
\begin{aligned}
F(r)=&\ur^2-\ur^{3-\ep}\uu^{\ep-1}\\
     &-(1-\ep)r\uu-5r^2.
\end{aligned}
\eeQ
By analyzing the monotonicity of $F(r)$, we obtain
\beQ
\begin{aligned}
F'''\geq 0&\Lrw F''(r)\leq F''\Big(1+\frac{tu}{2}\Big)=-(3-\ep)(2-\ep)2^{\ep-1}-8<0\\
&\Lrw F'(r)\leq F'(0)=0 \Lrw F(r)\leq F(0)=0.
\end{aligned}
\eeQ
This gives
\beQ
|(h_n-\bhn)(u,r)|\leq\frac{6x r\uu^{1-\ep}}{\ep(\ep-1)\ur^2}.
\eeQ
Therefore, (\ref{A4}) is proved by combining the above three cases.

Denote
\beq
k=\exp{\Bl-\frac{2\pi c^2x^2}{3}\Br} \leq 1. \label{kk}
\eeq
Then (\ref{3}) and (\ref{A4}) imply that
\beq
\begin{aligned}
\bgn(u,r) &\geq g_n(u,0)\\
          &=\exp{\Big[-4\pi \int_{0}^{\ift}{(h_n-\bhn)^2\frac{dr}{r}}\Big]}\\
          &\geq \exp{\Big[-\frac{2\pi c^2x^2}{3\uu^{2\ep}}\Big]}\geq k.
\end{aligned}\label{d}
\eeq

Next we prove the following claim.\\\\
{\em Claim II}:
\beq
(g_n -\bgn)(u,r)\leq \frac{4\pi c^2x^2}{3}\frac{r^2}{\uu^{2\ep-1}\ur^3}. \label{A6}
\eeq

Indeed, using (\ref{A4}) and the following equation
\beQ
\dfrac{\pt g_n}{\pr}=\frac{4\pi g_n(h_n-\bhn)^2}{r},
\eeQ
we obtain
\beQ
\begin{aligned}
(g_n -&\bgn)(u,r)\leq \frac{1}{r}\ir{\int_{r'}^{r}{\frac{\pt g_n}{\pt s}ds}dr'}\\
\leq &\frac{4\pi c^2x^2}{r\uu^{2\ep -2}}\Blb\frac{1}{2}\ir{\Big[\frac{1}{\Big(1+\frac{tu}{2}+r' \Big)^{2}}-\frac{1}{\ur^2}\Big]dr'}\\
     &-\frac{1+\frac{tu}{2}}{3}\ir{\Big[\frac{1}{\Big(1+\frac{tu}{2}+r' \Big)^3}-\frac{1}{\ur^3}\Big]dr'}\Brb\\
\leq & \frac{4\pi c^2x^2}{3}\frac{r^2}{\uu^{2\ep-1}\ur^3}.
\end{aligned}
\eeQ
Therefore the claim is proved.

Now (\ref{A1}) and (\ref{A6}) give
\beq
\begin{aligned}
\Ba-\frac{1}{2r}(g_n-\bgn)\bhn\Ba &\leq \frac{4\pi c^2x^3}{3\ep}  \frac{r}{\uu ^{3\ep -1}\ur ^{4}}\\
&\leq \frac{4\pi c^2x^3}{3\ep}\frac{1}{\uu^{2\ep+1}\ur^{1+\ep}}. \label{gnh}
\end{aligned}
\eeq

In the following we estimate $h_{n+1}$ and $\frac{\partial h_{n+1}}{\partial r} (u,r)$. These could be done by using
the characteristic $r(u)=\cn(u;r_1)$ through the line $r=r_1$ at $u=u_1$.

From (\ref{light-ray}), (\ref{d}), we have
\beq
r(u) = r_1 +\int_{u}^{u_1}{\frac{1}{2}\bgn du'}\geq r_1 + \frac{1}{2}k(u_1-u)\geq r_1 + \frac{1}{2}t k(u_1-u).\label{A7}
\eeq
Denote $r_0=r(0)$. Then (\ref{A7}) gives
\beq
|h(0,r_0)|\leq \frac{d}{(1+r_0)^{1+\ep}}\leq\frac{c_td}{\Big(1+\frac{tu_1}{2}+r_1 \Big)^{1+\ep}},\label{A12}
\eeq
\beq
\Ba\frac{\pt h}{\pr}(0,r_0)\Ba \leq \frac{d}{(1+r_0)^{1+\ep}} \leq\frac{c_t d}{\Big(1+\frac{tu_1}{2}+r_1 \Big)^{1+\ep}},\label{A18}
\eeq
A straightforward computation shows
\beQ
\int_{0}^{u_1}{\frac{du}{\uu^{2\ep+1}}}\leq a(u_1).
\eeQ
Using (\ref{A6}), (\ref{gnh}), we obtain
\beq
\begin{aligned}
\int_{0}^{u_1}{\Big[\frac{g_n-\bgn}{r}\Big]_{\cn}du}\leq \frac{4\pi c^2 x^2 a(u_1)}{3},\label{A11}
\end{aligned}
\eeq
and
\beq
\begin{aligned}
\int_{0}^{u_1}&{\Big[\Ba-\frac{1}{2r}(g_n-\bgn)\bhn\Ba\Big]_{\cn}du}\\
&\leq \frac{4\pi c^2x^3}{3\ep}\int_{0}^{u_1}{\frac{du}{\uu^{2\ep+1}\Big(1+\frac{u}{2}+r(u)\Big)^{1+\ep}}}\\
&\leq \frac{4\pi c^2x^3}{3\ep}\int_{0}^{u_1}{\frac{du}{\uu^{2\ep+1}\Big(1+\frac{u}{2}+r_1+\frac{tk}{2}(u_1-u)\Big)^{1+\ep}}}\\
&\leq \frac{4\pi c^2x^3c_t a(u_1)}{3\ep}\frac{1}{\Big(1+\frac{tu_1}{2}+r_1 \Big)^{1+\ep}}.
\label{A8}
\end{aligned}
\eeq
where $c_t$, $a(u_1)$ are given by (\ref{A9}), (\ref{A10}) respectively.

Now integrating (\ref{10}) along the characteristic $\cn$, we have
\beq
\begin{aligned}
h _{n+1}(u_1,& r_1)=h(0,r_0)\exp{\Big\{\int_{0}^{u_1}{\Big[\frac{g_n-\bgn}{2r}\Big]_{\cn}du}\Big\}}\\
&+\int_{0}^{u_1}\Big[-\frac{g_n-\bgn}{2r}\bhn\Big]_{\cn}
\exp\Big\{\int_{u}^{u_1}{\Big[\frac{g_n-\bgn}{2r}\Big]_{\cn}du}\Big\} du.
\label{a13}
\end{aligned}
\eeq
Substituting (\ref{A12}), (\ref{A11}) and (\ref{A8}) into (\ref{a13}), we obtain
\beq
\begin{aligned}
\Big(1+\frac{t u_1}{2}& +r_1 \Big)^{1+\ep} | h_{n+1}(u_1,r_1) | \\
&\leq  c_t\Bl d+\frac{4\pi c^2x^3a(u_1)}{3\ep}\Br   \exp{\Bl\frac{4\pi c^2x^2a(u_1)}{3}\Br}.
\end{aligned}\label{A13}
\eeq

Note that $\dfrac{\pt h_{n+1}}{\pr}$ satisfies the following equation (c.f. (9.16) in \cite{C1})
\beq
D_n\Bl\frac{\pt h_{n+1}}{\pr}\Br-\frac{1}{r}(g_n-\bgn)\frac{\pt h_{n+1}}{\pr}=f_n',\label{A14}
\eeq
where
\beQ
\begin{aligned}
f_n'=&-\frac{1}{2}\frac{\pt^2\bgn}{\prr}\bhn-\frac{1}{2}\frac{\pt\bgn}{\pr}\frac{h_n-\bhn}{r}
+\frac{1}{2}\frac{\pt^2\bgn}{\prr}h_{n+1},\\
\frac{\pt^2\bgn}{\prr}=&-\frac{2}{r^2}(g_n-\bgn)+\frac{4\pi}{r^2}(h_n-\bhn)^2g_n.
\end{aligned}
\eeQ
From (\ref{A1}) and (\ref{A6}), we have
\beq
\Big|\frac{\pt^2\bgn}{\prr}\Big|\leq \frac{8\pi c^2x^2}{\uu^{2\ep-1}\ur^3}.\label{A15}
\eeq
Using (\ref{A4}), (\ref{A1}), (\ref{A6}) and (\ref{A13}), we obtain
\beQ
\begin{aligned}
|f_n'|&\leq \frac{1}{2}\Big|\frac{\pt^2\bgn}{\prr}\Big||\bhn|+\frac{1}{2}\Big|\frac{\pt\bgn}{\pr}\Big|\Big|\frac{h_n-\bhn}{r}\Big|
+\frac{1}{2}\Big|\frac{\pt^2\bgn}{\prr}\Big||h_{n+1}|\\
&\leq \frac{\Big(\frac{8}{\ep}+2\Big)
c^3x^3+8\pi c^2x^2c_t\Big(d+\frac{4\pi c^2x^3a(u_1)}{3\ep}\Big)\exp\Big(\frac{4\pi c^2x^2a(u_1)}{3}\Big)}{\uu^{2\ep+1}\ur^{1+\ep}}.
\end{aligned}
\label{A16}
\eeQ

Using the same argument as in (\ref{A8}), we have
\beq
\int_{0}^{u_1}{\big[|f_n'|_{\cn}\big]du}\leq \frac{c_1c_ta(u_1)}{\Big(1+\frac{tu_1}{2}+r_1 \Big)^{1+\ep}},\label{A17}
\eeq
where
$$
c_1=\Big(\frac{8}{\ep}+2\Big) c^3x^3+c_t\frac{8\pi c^2x^2}{\ep}\Bl d+\frac{4\pi c^2x^3a(u_1)}{3\ep}\Br\exp{\Bl\frac{4\pi c^2x^2a(u_1)}{3}\Br}.
$$
Integrating (\ref{A14}) along the characteristic $\cn$, and using (\ref{A18}), (\ref{A17}), we obtain
\beq
\begin{aligned}
\Big(1+\frac{tu_1}{2}+& r_1\Big)^{1+\ep}\Ba\frac{\pt h_{n+1}}{\pr} (u_1, r_1)  \Ba \\
\leq & c_t\exp{\Bl\frac{4\pi c^2x^2a(u_1)}{3}\Br} \Big[d+\Big(\frac{8}{\ep}+2\Big) c^3x^3 a(u_1) \\
     &+8\pi c^2x^2c_ta(u_1)\Bl d+\frac{4\pi c^2x^3a(u_1)}{3\ep}\Br\exp{\Bl\frac{4\pi c^2x^2a(u_1)}{3}\Br}\Big].
\end{aligned}\label{A19}
\eeq
Thus (\ref{A13}) and (\ref{A19}) indicate
\beQ
\|h_{n+1}\|_{X_{u_1,t}} \leq F_{t}(u_1,x).
\eeQ
Therefore the proof of the lemma is complete. \qed

\begin{lem}\label{lemB}
Let initial data $\breve{h}(r)$ satisfy (\ref{h0}). If there exists some constant $x>0$ such that
\beQ
\|h_n(u,r)\|_{X_{u_1,t}}\leq x,
\eeQ
then it satisfies that
\beq
\|(h_{n+1}-h_n)(u,r) \|_{Y_{u_1,t}} \leq C_{t}(u_1,x) \|(h_n-h_{n-1})(u,r)\|_{Y_{u_1,t}}. \label{h2-h1}
\eeq
\end{lem}
\pf Note that the following equation holds (c.f. (9.27) in \cite{C1})
\beq
\begin{aligned}
D_n(h_{n+1}-h_n)-&\frac{1}{2}\frac{\pt \bgn}{\pr}(h_{n+1}-h_n)\\
                =&\frac{1}{2}(\bgn -\bg_{n-1})\frac{\phn}{\pr}+\frac{1}{2}\Bl\frac{\pt\bgn}{\pr}-\frac{\pt\bg_{n-1}}{\pr}\Br h_n\\
                 &+f_n-f_{n-1},
\end{aligned}\label{B1}
\eeq
where
\beQ
f_n-f_{n-1}=-\frac{1}{2}\frac{\pt\bgn}{\pr}(\bhn-\bh_{n-1})-\frac{1}{2}\Bl\frac{\pt\bgn}{\pr}-\frac{\pt\bg_{n-1}}{\pr}\Br \bh_{n-1}.
\eeQ
Now (\ref{A2}) gives
\beq
|h_n-\bhn+h_{n-1}-\bh_{n-1}|\leq \frac{2cxr}{\uu^{\ep-1}\ur^2}.\label{B2}
\eeq
On the other hand, using the definition, we can prove
\beq
|h_n-\bhn-h_{n-1}+\bh_{n-1}|\leq \frac{4\|h_n-h_{n-1}\|_{Y_{u_1,t}}}{\ep\uu^{\ep}\ur}.\label{B3}
\eeq
Multiplying (\ref{B2}) by (\ref{B3}), we have
\beq
|(h_n-\bhn)^2-(h_{n-1}-\bh_{n-1})^2|\leq \frac{8cxr\|h_n-h_{n-1}\|_{Y_{u_1,t}}}{\ep\uu^{2\ep-1}\ur^3}.\label{B4}
\eeq
As
\beQ
x_1,\, x_2<0 \Longrightarrow |e^{x_1}-e^{x_2}|\leq |x_1-x_2|,
\eeQ
we obtain
\beq
\begin{aligned}
|g_n-g_{n-1}|&\leq 4\pi\Ba\int_{r}^{\ift}{\Big[(h_n-\bhn)^2-(h_{n-1}-\bh_{n-1})^2\Big]\frac{dr'}{r'}\Ba}\\
&\leq \frac{32\pi c x}{\ep\uu^{2\ep-1}} \Big[\int_{r}^{\ift}{\frac{dr'}{\Big(1+\frac{t u}{2}+r' \Big)^3}}\Big] \|h_n-h_{n-1}\|_{Y_{u_1,t}} \\
&\leq \frac{16\pi c x }{\ep\uu^{2\ep-1}\ur^2}  \|h_n-h_{n-1}\|_{Y_{u_1,t}}.
\label{B5}
\end{aligned}
\eeq
This implies
\beq
|\bgn-\bg_{n-1}|\leq \frac{16\pi c x}{\ep\uu^{2\ep}\ur} \|h_n-h_{n-1}\|_{Y_{u_1,t}}. \label{B6}
\eeq

By (\ref{A4}), (\ref{B4}) and (\ref{B5}), we obtain
\beQ
\begin{aligned}
\Ba\frac{\pt(g_n-g_{n-1})}{\pr}\Ba
=&\Ba\frac{4\pi(h_n-\bhn)^2g_n}{r}-\frac{4\pi(h_{n-1}-\bh_{n-1})^2g_{n-1}}{r}\Ba\\
\leq & \frac{4\pi}{r}\Big[|g_n||(h_n-\bhn)^2-(h_{n-1}-\bh_{n-1})^2 |\\
&+|g_n-g_{n-1}||h_{n-1}-\bh_{n-1}|^2 \Big]\\
\leq &\frac{8\pi (4cx+8\pi c^3x^3) }{\ep\uu^{2\ep-1}\ur^3}\|h_n-h_{n-1}\|_{Y_{u_1,t}}.
\end{aligned}
\eeQ
Thus
\beq
\begin{aligned}
\Ba\frac{\pt(\bgn-\bg_{n-1})}{\pr}\Ba&\leq\frac{1}{r^2}\ir{\int_{r'}^{r}{\frac{\pt(g_n-g_{n-1})}{\pt s}ds}dr'}\\
&\leq\frac{4\pi (4cx+8\pi c^3x^3)}{\ep\uu^{2\ep+1}\ur^{1+\ep}} \|h_n-h_{n-1}\|_{Y_{u_1,t}}.
\end{aligned}\label{B7}
\eeq
Using (\ref{A6}), we have
\beq
\begin{aligned}
|f_n-f_{n-1}|&\leq \frac{1}{2}\Ba\frac{\pt\bgn}{\pr}\Ba|\bhn-\bh_{n-1}|+\frac{1}{2}|\bh_{n-1}|\Ba\frac{\pt\bgn}{\pr}-\frac{\pt\bg_{n-1}}{\pr}\Ba\\
&\leq \frac{4\pi c^2 x \Big(\frac{1}{\ep}+\frac{1}{\ep ^{2}}\Big)
(6cx+8\pi c^3x^3)}{\uu^{3\ep-1}\ur^3 }\|h_n-h_{n-1}\|_{Y_{u_1,t}}.
\label{B8}
\end{aligned}
\eeq
From (\ref{B6}), (\ref{B7}) and the assumption on $h_n$, we obtain
\beq
|\bgn-\bg_{n-1}|\Ba\frac{\phn}{\pr}\Ba\leq\frac{16\pi cx^2}{\ep\uu^{2\ep}\ur^{2+\ep}}
\|h_n-h_{n-1}\|_{Y_{u_1,t}},\label{B9}
\eeq
\beq
\Ba\frac{\pt(\bgn-\bg_{n-1})}{\pr}\Ba |h_n| \leq\frac{4\pi x (4cx+8\pi c^3x^3) }{\ep\uu^{2\ep}\ur^{3+\ep}}
\|h_n-h_{n-1}\|_{Y_{u_1,t}}.\label{B10}
\eeq
Now applying (\ref{B8}), (\ref{B9}) and (\ref{B10}) to (\ref{B1}), we can prove
\beQ
\begin{aligned}
\Ba D_n(h_{n+1}&-h_n)-\frac{1}{2}\frac{\pt\bgn}{\pr}(h_{n+1}-h_n)\Ba\\
&\leq
\frac{32\pi \Big(\frac{1}{\ep}+\frac{1}{\ep ^{2}}\Big)(3+4\pi c^2 x^2)c^3 x^2}{\uu^{2\ep+1}\ur^{1+\ep}} \|h_n-h_{n-1}\|_{Y_{u_1,t}}.
\end{aligned}
\eeQ
Integrating (\ref{B1}) along the characteristic $\cn$ and using the initial condition
\beQ
(h_{n+1}-h_n)(0,r)=0,
\eeQ
we obtain
\beQ
\Big(1+\frac{t u_1}{2}+r_1 \Big)^{1+\ep} \Big|(h_{n+1}-h_n)(u_1,r_1)\Big| \leq  C_{t}(u_1,x) \|h_n-h_{n-1}\|_{Y_{u_1,t}}.
\eeQ
Thus,
\beQ
\|(h_{n+1}-h_n)\|_{Y_{u_1,t}}\leq C_{t}(u_1, x) \|h_n-h_{n-1}\|_{Y_{u_1,t}}.
\eeQ
Therefore the proof of the lemma is complete.\qed

\begin{lem}\label{lemC}
If $t=0$ and $x>2d$, there exists $\gamma>0$ such that for any
\beQ
u_0 \in (0, \gamma ],
\eeQ
the sequence $\{h_n\}$ is uniformly bounded by $x$ in $X_{u_0,0}$ and contracts in $Y_{u_0,0}$. Moreover, both $\{h_n\}$ and $\Big\{\dfrac{\phn}{\pr}\Big\}$are equibounded and equicontinuous in $I_0=[0,u_0]\times[0,\ift)$.
\end{lem}
\pf Note that, in inequalities (\ref{A20}) and (\ref{h2-h1}), $F_{0}(u,x)$ and $C_{0}(u,x)$ are strictly monotonically increasing with respect to $u<\infty$, and
$$
F_{0}(0,x)=2d<x, \quad F_{0}(\infty, x)=\infty, \quad C_{0}(0,x)=0,\quad C_{0}(\infty,x)=\infty.
$$
There are $\gamma_1 >0$, $\gamma_2 >0$ such that
\beQ
F_{0}(\gamma_1,x)=x, \quad C_{0}(\gamma_2,x)=\frac{1}{2}.
\eeQ
Then (\ref{A20}) and (\ref{h2-h1}) imply that, for any
\beQ
u_0 \in (0, \gamma], \quad \gamma=\min\big\{\gamma_1, \gamma_2\big\},
\eeQ
the following inequalities hold
\beQ
\|h_{n}\|_{X_{u_0,0}} \leq x \Longrightarrow
\begin{matrix}
\left\{
\begin{aligned}
& \|h_{n+1}\|_{X_{u_0,0}} \leq x,\\
& \|h_{n+1}-h_{n}\|_{Y_{u_0,0}} \leq \frac{1}{2} \|h_{n}-h_{n-1}\|_{Y_{u_0,0}}.
\end{aligned}
\right.
\end{matrix}
\eeQ
This also indicates that $\{h_n\}$ is equibounded in $I_0$. Since
\beQ
\Ba\frac{\pt h_{n+1}}{\pr}(u,r)\Ba\leq\frac{x}{(1+r)^{1+\ep}},
\eeQ
we have
\beq
\begin{aligned}
\Ba\frac{\pt h_{n+1}}{\pu}(u,r)\Ba \leq& \frac{1}{2}|\bar g_n|\Ba\frac{\pt h_{n+1}}{\pr}\Ba+\frac{1}{2r}|g_n-\bgn||h_{n+1}|+\frac{1}{2r}|g_n-\bgn||\bhn|\\
\leq &\frac{1}{2}\Bl 1+\frac{4\pi c^2x^2}{3}+\frac{8\pi c^2x^2}{3\ep}\Br\frac{x}{(1+r)^{1+\ep}}.
\end{aligned}\label{46}
\eeq
Therefore $\Big\{\dfrac{\pt h_{n+1}}{\pr}\Big\}$ and $\Big\{\dfrac{\pt h_{n+1}}{\pu}\Big\}$ are equibounded in $I_0$, which implies that $\{h_n\}$ is equicontinuous in $I_0$.

In the following we show that $\Big\{\dfrac{\pt h_n}{\pr}\Big\}$ is equicontinuous in $I_0$. For any
\beQ
u_1\in [0,u_0], \quad 0\leq r_1<r_2,
\eeQ
denote $\cn(u;r_1)$ and $\cn(u;r_2)$ the two characteristics through the line $u=u_1$ at $r=r_1$ and $r=r_2$ respectively. Denote
\beq
k'=\exp{\Big(\frac{4\pi c^2x^2}{3k}\Big)}, \label{k}
\eeq
where $k$ is given by (\ref{kk}). From (\ref{d}) and (4.29) of \cite{C2}, we obtain
\beQ
\begin{aligned}
\cn(u;r_2)-\cn(u;r_1)&\leq(r_2-r_1)\sup_{s\in[r_1,r_2]}\exp\Big\{\frac{1}{2}\int_{u_1}^{u}{\Big[\frac{\pt \bgn}{\pr}\Big] _{\cn(u';s)}du'}\Big\}\\
&\leq k' (r_2-r_1).
\end{aligned}
\eeQ
Therefore, if we denote
\beQ
\Theta(f)(u)=f\big(u,\cn(u;r_1)\big)-f\big(u,\cn(u;r_2)\big)
\eeQ
for any differentiable function $f$, we can find that
\beQ
\begin{aligned}
|\Theta(f)(u)| & \leq \sup {\Ba \frac{\pt f}{\pr}\Ba} \big|\cn(u;r_1)-\cn(u;r_2) \big|  \\
               & \leq k' \sup{\Ba \frac{\pt f}{\pr}\Ba}(r_2-r_1).
\end{aligned}
\eeQ

Define
\beq
\psi(u)=\frac{\pt h_{n+1}}{\pr}\big(u,\cn(u;r_1)\big)-\frac{\pt h_{n+1}}{\pr}\big(u,\cn(u;r_2)\big). \label{47}
\eeq
Differentiating (\ref{47}) and using (\ref{A14}), we obtain
\beq
\psi'(u)-\frac{\big(g_n-\bgn \big)\big(u,\cn(u;r_1)\big)}{\cn(u;r_1)}\psi(u)=\sum _{i=1} ^4 A_i,\label{psiU}
\eeq
where
\beQ
\begin{aligned}
A_1&=\frac{\pt h_{n+1}}{\pr}(u,\cn(u;r_2))\Theta\Bl\frac{g_n-\bgn}{r}\Br(u),\\
A_2&=\frac{1}{2}\Theta\Bl\frac{\pt^2\bgn}{\pt r^2}(h_{n+1}-\bar{h}_{n+1})\Br(u),\\
A_3&=\frac{1}{2}\Theta\Bl\frac{\pt^2\bgn}{\pt r^2}(\bar{h}_{n+1}-\bhn)\Br(u),\\
A_4&=-\frac{1}{2}\Theta\Bl\frac{\pt\bgn}{\pr}\frac{h_n-\bhn}{r}\Br(u).
\end{aligned}
\eeQ

Now we estimate $A_i$ for $1\leq i \leq 4$. Using (\ref{A20}), (\ref{A15}) and
\beQ
\dfrac{g_n-\bgn}{r}=\dfrac{\pt\bgn}{\pr},
\eeQ
we obtain
\beQ
\begin{aligned}
|A_1|&\leq \frac{x}{\Big(1+\cn(u;r_2)\Big)^{1+\ep}} \Ba\frac{\pt^2\bgn}{\prr}\Ba k' |r_2-r_1| \\
&\leq \frac{8\pi c^2x^3 k'}{\Big(1+\cn(u;r_2)\Big)^{1+\ep} (1+r)^3}|r_2-r_1| \\
&\leq 8\pi c^2x^3 k' |r_2-r_1|.
\end{aligned}
\eeQ

To estimate $A_2$ we need to estimate the 3rd partial derivative of $\bar g_n$ with respect to $r$. It is straightforward that
\beq
\begin{aligned}
\frac{\partial ^3 \bar g_n}{\partial r^3}=&\frac{6}{r^3}(g_n -\bar g _n )-\frac{16\pi}{r^3}(h_n -\bar h _n)^2 g_n \\
& +\frac{8\pi}{r^2}(h_n -\bar h _n)\frac{\partial (h_n -\bar h _n)}{\partial r}g_n\\
& +\frac{16\pi ^2}{r^3} (h_n -\bar h _n )^4 g_n.
\end{aligned} \label{3rd}
\eeq
Then
\beq
\Big|\frac{\partial ^3 \bar g_n}{\partial r^3}\Big|\leq \frac{1}{r}(40\pi c^2x^2+16\pi^2c^4x^4). \label{3rd-1}
\eeq
Thus, (\ref{A4}), (\ref{3rd}), (\ref{3rd-1}) give
\beQ
\begin{aligned}
|A_2| \leq &\frac{k'}{2}\Big[\Big|\frac{\partial ^3 \bar g_n}{\partial r^3}\Big||h_{n+1}-\bar{h}_{n+1}|\\
           & +\Big|\frac{\partial ^2 \bar g_n}{\partial r^2}\Big|\Big|\frac{\pt (h_{n+1}-\bar{h}_{n+1})}{\pt r}\Big|\Big]|r_2-r_1|\\
      \leq & (28\pi c^3x^3+8\pi^2c^5x^5)k'|r_2-r_1|.
\end{aligned}
\eeQ

Now we use Lemma \ref{lemB} to estimate $A_3$. Denote
\beQ
I_1 =[0,u_0]\times [0,1].
\eeQ
As $C_0(u_0,x)\leq \frac{1}{2}$ in Lemma \ref{lemB}, we know that
\beQ
h_{n+1}-h_n \lrw 0
\eeQ
uniformly. Using (\ref{A15}), we obtain
$$\frac{\pt^2\bgn}{\pt r^2}(h_{n+1}-h_n) \lrw 0$$
uniformly. This leads to
$$
\lim\limits_{n\rw+\infty}{\sup\limits_{(u,r)\in I_1}{\Big|\frac{\pt^2\bgn}{\pt r^2}(\bar h_{n+1}-\bar h_n)\Big|}}\leq \lim\limits_{n\rw+\infty}{\sup\limits_{(u,r)\in I_1}{\Big|\frac{\pt^2\bgn}{\pt r^2}(h_{n+1}-h_n)\Big|}}=0.
$$
This implies that
$$\frac{\pt^2\bgn}{\pt r^2}(\bar h_{n+1}-\bar h_n) \lrw 0$$
uniformly. Therefore the sequence
$$\Big\{\frac{\pt^2\bgn}{\pt r^2}(\bar h_{n+1}-\bar h_n)\Big\}$$
is equicontinuous in $I_1$, and, for any $\eta>0$, there exists $t_1$ such that
$$|r_2'-r_1'|=|\cn(u;r_1)-\cn(u;r_2)|\leq t_1 \Longrightarrow |A_3|\leq \frac{\exp(-4\pi c^2x^2)}{3 u_0} \eta.$$
The same argument gives
\beQ
|A_3|\leq (48\pi c^3x^3+16\pi^2c^4x^5)k'|r_2-r_1|
\eeQ
for $(u,r)\in [0,u_0]\times [1,\infty)$. Take
\beQ
s_1=\min{\Big\{\frac{t_1}{k'},\frac{\exp(-4\pi c^2x^2)}{3k'(48\pi c^3x^3+16\pi^2c^4x^5)u_0} \eta   \Big\}},
\eeQ
then for any $(u,r_1),\,(u,r_2)\in I_0$,
\beQ
|r_2-r_1|\leq s_1 \Lrw |r_2 '-r_1 '|\leq k's_1\leq t_1 \Lrw |A_3|\leq \frac{\exp(-4\pi c^2x^2)}{3 u_0} \eta.
\eeQ

Finally, we estimate $A_4$. As
\beQ
\begin{aligned}
\frac{\partial}{\partial r} \Bl\frac{\pt\bgn}{\pr}\frac{h_n-\bhn}{r}\Br=&\frac{\pt ^2\bgn}{\pr ^2}\frac{h_n-\bhn}{r}
+\frac{1}{r}\frac{\pt\bgn}{\pr}\frac{\partial (h_n-\bhn)}{\partial r}\\
& -\frac{1}{r^2} \frac{\pt\bgn}{\pr}(h_n-\bhn),
\end{aligned}
\eeQ
it is bounded in $I_0$ and we find
\beQ
|A_4|\leq 6\pi c^3x^3k'|r_2-r_1|.
\eeQ
On the other hand,
\beQ
\Ba\frac{(g_n-\bgn)(u,\cn(u;r_1))}{\cn(u;r_1)}\Ba\leq 4\pi c^2x^2.
\eeQ
Integrating (\ref{psiU}), we obtain
\beQ
\begin{aligned}
\psi(u_1)=&\psi(0)\exp{\Big[\int_{0}^{u_1}{\frac{\big(g_n-\bgn \big)\big(u,\cn(u;r_1)\big)}{\cn(u;r_1)}du}\Big]}\\
          &+\int_{0}^{u_1}\Big\{\exp\Big[\int_{u}^{u_1}{\frac{\big(g_n-\bgn \big)\big(u',\cn(u';r_1)\big)}{\cn(u';r_1)}du' \Big]}\sum _{i=1} ^4 A_i \Big\}du .
\end{aligned}
\eeQ
Therefore, the above estimates indicate that if $|r_2-r_1|\leq s_1$ then
$$|\psi(u_1)|\leq \exp(4\pi c^2x^2)\Big[|\psi(0)|+(42\pi c^3x^3+8\pi^2 c^5x^5)  k' u_0 (r_2-r_1)\Big]+\frac{\eta}{3}.$$

Now we estimate $\psi(0)$. In $I_0$, we define
$$\omega(t)=\sup_{|r_1''-r_2''|\leq t}{\Blb\Ba\frac{\pt h}{\pr}(0,\cn(0;r_1))-\frac{\pt h}{\pr}(0,\cn(0;r_2))\Ba\Brb},$$
where
\beQ
r_1 ''=\cn(0;r_1), \quad r_2 ''=\cn(0;r_2).
\eeQ
From the basic analysis, we know that $\dfrac{\ph}{\pr}(0,r)$ is uniformly continuous if and only if
\beQ
\lim\limits_{t\rw 0}{\omega(t)}=0.
\eeQ
Thus, for any $\eta>0$, there exists $t_2>0$ such that, for any $t\leq t_2$,
$$\omega(t)<\frac{\exp{(-4\pi c^2x^2)}}{3}   \eta.$$
Denote
\beQ
s_2=\frac{\exp{(-4\pi c^2x^2)}}{3(42\pi c^3x^3+8\pi^2 c^5x^5) k' u_0}.
\eeQ
Choosing
\beQ
s=\min{\Big\{s_1,\frac{t_2}{k'},s_2\Big\}},
\eeQ
we find that
$$
|r_2-r_1|\leq s \Lrw
\begin{matrix}
\left\{
\begin{aligned}
& |r_1 '-r_2 '|\leq t_1,\\
& |r_1''-r_2''|\leq k' s\leq t_2
\end{aligned}
\right.
\end{matrix}
\Lrw |\psi(u_1)|\leq \eta.
$$
Thus we have
\beQ
|\psi(u_1)|\leq \eta .
\eeQ
This implies that
\beQ
\Ba\frac{\pt h_{n+1}}{\pr}(u_1,r_1)-\frac{\pt h_{n+1}}{\pr}(u_1,r_2)\Ba\leq\eta.
\eeQ
Hence $\Big\{\dfrac{\pt h_{n+1}}{\pr}\Big\}$ is equicontinuous with respect to $r$ in $I_0$.

The equicontinuous of $\Big\{\dfrac{\pt h_{n+1}}{\pr}\Big\}$ with respect to $u$ can be proved by the equiboundedness of $D_n\Bl\frac{\pt h_{n+1}}{\pr}\Br.$ Therefore the proof of the lemma is complete.\qed

\mysection{Local existence and uniqueness of classical solutions}\ls

In this section we prove local existence and uniqueness of classical solutions.

\begin{thm}
For any initial data $\breve{h}(r) \in C^1[0,\ift)$ which satisfies (\ref{wave-like}), then there exists $\uz>0$ and a unique classical solution
\beQ
h(u,r) \in C^1 \Big([0,\uz]\times [0,\ift)\Big)
\eeQ
of (\ref{3}) which satisfies the initial condition $h(0,r)=\breve{h}(r)$ and the decay property (\ref{wave-like}) uniformly in $u\in [0, u_0]$.
\end{thm}
\pf We use the main estimates for the case $t=0$ derived in lemmas in the last section. Let $u_0<\gamma$
in Lemma \ref{lemC}. By Arzela-Ascoli theorem, there exists a subsequence $\{\hni\}$ and function $\hat{h}$ such that
\beq
\hni \lrw \hat{h},\quad \dfrac{\pt\hni}{\pr} \lrw \dfrac{\pt \hat{h}}{\pr} \label{hni-hath}
\eeq
uniformly in any compact subset $[0,u_0]\times [0,r_0]$. Moreover, $\hat{h} \in X_{u_0, 0}$, and
\beq
\|\hat{h}(u,r)\|_{X_{u_0, 0}}\leq x.          \label{58}
\eeq
Therefore the convergence of (\ref{hni-hath}) is uniform in $I_0$, and $\Big\{\dfrac{\pt \hat{h}}{\pr}\Big\}$ is equibounded in $I_0$.

By the contraction principle,
\beQ
h_n \lrw h
\eeQ
in $Y_{u_0, 0}$. Therefore
\beQ
h=\hat{h} \in X_{u_0, 0}.
\eeQ

Using (\ref{A4}), (\ref{A1}) and the arguments in the proofs of Lemma \ref{lemA}, Lemma \ref{lemC}, we can deduce that
\beQ
\bhi \lrw \bh, \quad
\ani =\int_{r}^{\ift}{(\hni-\bhi)^2\dfrac{dr'}{r'}} \lrw  A =\int_{r}^{\ift}{(h-\bh)^2\dfrac{dr'}{r'}}
\eeQ
uniformly in $I_0$, which implies that
\beQ
\gni=\exp{(-4\pi \ani)}\lrw g=\exp{(-4\pi A)}, \quad \bgi \lrw \bar g
\eeQ
uniformly in $I_0$, and
\beQ
\dfrac{\pt\bhi}{\pr} \lrw \dfrac{\pt \bh}{\pr}, \quad \dfrac{\pt\gni}{\pr} \lrw \dfrac{\pt g}{\pr}, \quad
\dfrac{\pt\bgi}{\pr} \lrw \dfrac{\pt \bar g}{\pr}
\eeQ
uniformly in $I_0$.

Let $\cni(u;r_1)$, $\chi(u;r_1)$ be the characteristics corresponding to $h_{n_i}$, $h$ through $r=r_1$ at $u=u_1$.
Then
\beQ
\dfrac{d(\cni-\chi)}{du}=-\dfrac{1}{2}\Big[\bgi(u,\cni)- \bg(u,\chi)\Big].
\eeQ
And it satisfies the initial condition
\beQ
(\cni-\chi)(u_1)=\cni(u_1, r_1)-\chi(u_1,r_1)=r_1-r_1=0.
\eeQ
We obtain
\beQ
\begin{aligned}
(\cni-\chi)(u)=&-\frac{1}{2}\int ^u _{u_1} \Big[\bgi(u',\cni)- \bg(u',\chi)\Big]du'\\
              =&-\frac{1}{2}\int ^u _{u_1} \Big[\bgi(u',\cni)- \bgi(u',\chi)\Big]du'\\
               & -\frac{1}{2}\int ^u _{u_1} \Big[\bgi(u',\chi)- \bg(u',\chi)\Big]du'.
\end{aligned}
\eeQ
Thus, using the convergence of $\bgi(u',\chi)$ to $\bg(u',\chi)$, we can show
\beq
\cni(u;r_1)\lrw\chi(u;r_1) \label{chi-1}
\eeq
uniformly in $I_0$.

Applying (\ref{a13}) to $h_{n_i}$, taking $i \rw \infty$ and using (\ref{A6}), (\ref{gnh}), (\ref{chi-1}), we obtain
\beq
\begin{aligned}
h(u_1,r_1)=&h(0,\chi(0;r_1))\exp\Big\{\int_{0}^{u_1}{\Big[\frac{g-\bg}{2r}\Big]_{\chi}}du\Big\}\\
&+\int_{0}^{u_1}{\Big[-\frac{g-\bg}{2r}\bh\Big]_{\chi} \exp\Big\{\int_{u}^{u_1} \Big[\frac{g-\bg}{2r}\Big]_{\chi}du' \Big\} }du
\label{68}
\end{aligned}
\eeq
in $I_0$. This implies that $h$ satisfies (\ref{3}) in the integral sense in $I_0$. As $\frac{\partial h}{\partial r}$ is continuous in $I_0$, it follows that $h$ is continuously differential with respect to $u$ in $I$. Therefore, $h$ satisfies (\ref{3}) in the differential sense in $I_0$ and satisfies the initial condition.

Now we prove the uniqueness. Similar to \cite{C1}, we define
\beQ
\Delta(u)=\sup_{r\geq 0}{\Big\{(1+r)^{1+\ep}|h_1-h_2|\Big\}}.
\eeQ
Let $D_1$, $D_2$ are the $D$-operator corresponding to the solutions $h_1$, $h_2$ respectively. We have
\beQ
\begin{aligned}
D_1(h_1-h_2)=& D_1 h_1- D_2 h_2 + \frac{1}{2}(\bg_1-\bg_2)\frac{\pt h_2}{\pr}\\
=&\frac{1}{2r}(g_1-\bg_1-g_2+\bg_2)(h_2-\bh_2)\\
 &+\frac{1}{2r}(g_1-\bg_1)(h_1-\bh_1-h_2+\bh_2)\\
 &+\frac{1}{2}(\bg_1-\bg_2)\frac{\pt h_2}{\pr}.
\end{aligned}
\eeQ

From the proofs of Lemma \ref{lemA}, Lemma \ref{lemB}, we can derive, for $i=1$, $2$,
\beQ
|(h_i-\bh_i)(u,r)|\leq \frac{cc_0r}{(1+r)^2}, \quad
|(g_i-\bg_i)(u,r)|\leq \frac{4\pi c^2c_0^2r^2}{3(1+r)^3}.
\eeQ
Therefore we obtain
\beQ
\begin{aligned}
|h_1-\bh_1+h_2-\bh_2| & \leq |h_1-\bh_1|+|h_2-\bh_2| \leq \frac{2cc_0r}{(1+r)^2},\\
|h_1-\bh_1-h_2+\bh_2| & \leq \frac{\Delta}{(1+r)^{1+\ep}}+\frac{2\Delta}{\ep(1+r)} \leq\frac{3\Delta}{\ep(1+r)}.
\end{aligned}
\eeQ
Multiply them, we obtain
\beQ
|(h_1-\bh_1)^2-(h_2-\bh_2)^2|\leq \frac{6cc_0\Delta r}{\ep (1+r)^3}.
\eeQ
Therefore
\beQ
\begin{aligned}
|g_1-g_2|&\leq 4\pi \Ba\int_{r}^{\ift}{(h_1-\bh_1)^2\frac{dr'}{r'}}-\int_{r}^{\ift}{(h_2-\bh_2)^2\frac{dr'}{r'}}\Ba\\
&\leq \frac{24\pi cc_0\Delta}{\ep}\int_{r}^{\ift}{\frac{dr}{(1+r)^3}}\leq \frac{12\pi cc_0\Delta}{\ep (1+r)^2}.
\end{aligned}
\eeQ
And
\beQ
|\bg_1-\bg_2|\leq\frac{1}{r}\ir{|g_1-g_2|dr}\leq \frac{12\pi cc_0\Delta}{\ep(1+r)}.
\eeQ
The above two inequality give
\beQ
|g_1-\bg_1-g_2-\bg_2|\leq |g_1-g_2|+|\bg_1-\bg_2|\leq \frac{24\pi cc_0\Delta}{\ep(1+r)}.
\eeQ
Finally, we obtain
\beQ
|D_1(h_1-h_2)|\leq \frac{20\pi c^2c_0^2\Delta}{\ep(1+r)^{1+\ep}}.
\eeQ
Integrating it along the characteristic $\chi_1$ which intersects the line $u=u_1$ at $r=r_1$ and using the initial condition
\beQ
(h_1-h_2)(0, r)=0,
\eeQ
we obtain
\beQ
\Delta(u_1)\leq \frac{20\pi c^2c_0^2}{\ep}\int_{0}^{u_1}{\Delta(u)du}.
\eeQ
By the Gronwall inequality, we have
\beQ
\Delta(u)=0
\eeQ
for any $u \leq u_1 \in [0,u_0]$. Thus the uniqueness holds and the proof of the theorem is complete. \qed

\mysection{Global existence and uniqueness for small initial data}

In this section, we prove the global existence and uniqueness for small initial data.

\begin{thm}\label{Glo}
Consider initial data $\breve{h}(r)\in C^1[0,\ift)$ which satisfies (\ref{wave-like}). Denote
\beQ
d_0 : =\inf_{b>0}\sup_{r\geq0}{\Blb \Big(1+\frac{r}{b}\Big)^{1+\ep}\Bl|h_0(r)|+\Ba b\frac{\pt h_0}{\pr}(r)\Ba\Br\Brb}.
\eeQ
Then there exists $\delta >0$ such that if $d_0 <\delta$, there exists a unique global classical solution
\beQ
h(u,r) \in C^1 \Big([0,\infty)\times [0,\ift)\Big)
\eeQ
of (\ref{3}) which satisfies the initial condition $h(0,r)=\breve{h}(r)$ and the decay property
\beQ
|h(u,r)|\leq \frac{C}{\Big(1+\frac{u}{2}+r \Big)^{1+\ep}},\quad \Ba\frac{\pt h}{\pr}(u,r)\Ba\leq\frac{C}{\Big(1+\frac{u}{2}+r \Big)^{1+\ep}}
\eeQ
for some constant $C$ depending on $\ep$ only. Moreover, the corresponding spacetime is future causally geodesically complete with vanishing final Bondi mass $M_{B1}$.
\end{thm}
\pf Let $u_1=\infty$, $t=1$ in Lemma \ref{lemA} and Lemma \ref{lemB}. It is clearly that $h_0 \in X_{0,1}$ where $h_0$ given by (\ref{h_0}). Suppose
\beQ
\|h_0\|_{X_{0,1}}=d.
\eeQ
Let $\{h_n \}$ be the sequences constructed by (\ref{10}) and $h_n$ satisfies
\beQ
\|h_n\|_{X_{\infty,1}}\leq x
\eeQ
for some $x>0$. Then, by (\ref{A20}), we have
\beQ
\|h_{n+1}\|_{X_{\infty,1}}\leq F_1 (\infty, x).
\eeQ

Now we define a function
\beQ
G(x)=\frac{x}{2+\frac{8\pi c^2x^2}{\ep}} \exp\Big(-\frac{4\ep+4 \ep ^2+8}{3\ep}\pi c^2 x^2\Big)-\frac{24+6\ep+4\pi}{3\ep^2} c^3x^3
\eeQ
for $x \geq 0$. It is easy to see that
\beQ
G(x) \leq \frac{x}{2},
\eeQ
and
\beQ
F_1 (\infty, x)\leq x \Longleftrightarrow G(x)\geq d.
\eeQ
The direct computation shows that
\beQ
G(0)=0, \quad G'(0)=\frac{1}{2}, \quad G(\infty)=-\infty, \quad G'(\infty)=-\infty.
\eeQ
Therefore there exists some $x_1>0$ such that $G(x)$ attains its maximum value $G_0>0$ at $x_1>0$, and $G(x)$ is strictly monotonically increasing in $[0, x_1]$.

Choose $d<G_0$. Then there exists $x_0 >0$ such that
\beQ
d=G(x_0) <\frac{x_0}{2}.
\eeQ
Thus we obtain a sequence $\{h_n \}$ such that, for any $x \in [x_0, x_1]$,
\beQ
\|h_n\|_{X_{\infty,1}}\leq x.
\eeQ
For this $x$, Lemma \ref{lemB} gives
\beQ
\|(h_{n+1}-h_n)\|_{Y_{\infty,1}}\leq C_{1}(\infty, x) \|h_n-h_{n-1}\|_{Y_{\infty,1}},
\eeQ
where
\beQ
C_{1}(\infty, x)=\frac{32\pi (\ep +1)}{\ep ^{3}}  c^3 x^2 (3+4\pi c x ) \exp\Big({\frac{2\ep +2\ep ^2 +4}{3\ep}}\pi c^2 x^2 \Big).
\eeQ
As $C_{1}(\infty, x)$ is strictly monotonically increasing for $x \geq 0$ and $C_{1}(\infty, 0)=0$, we can find $x_2>0$ such that
\beQ
C_{1}(\infty, x_2)=\frac{1}{2}.
\eeQ
Thus, for any $x\in (0,x_2)$,
\beQ
\|h_{n+1}-h_n\|_{Y_{\infty,1}}<\frac{1}{2}\|h_n-h_{n-1}\|_{Y_{\infty,1}}.
\eeQ

Now define
\beQ
\delta= \max_{x\in[0, x_2]}{G(x)}>0.
\eeQ
If $d<\delta$, then we can find $x<x_2$ such that $$G(x)\geq d.$$ Then using the same argument as the proof of Theorem \ref{thm1}, we can
find $h$ such that
\beQ
h_n \lrw h, \quad \frac{\partial h_n}{\partial r} \lrw \frac{\partial h}{\partial r}
\eeQ
uniformly in $I_0$. Moreover, $h$ is a global classical solution of (\ref{3}).

Now let $h(0,r)=\breve{h}(r)$ be the initial data $\breve{h}(r)$ with $d_0<\delta$. Then there exists $b>0$ such that
\beQ
\sup_{r\geq 0}{\Blb\Big(1+\frac{r}{b}\Big)^{1+\ep}\Bl|\breve{h}(r)|+\Ba b\frac{\pt \breve{h}}{\pr}(r)\Ba\Br\Brb}<\delta.
\eeQ
Define new initial data
\beQ
\tilde{h}(0,r)=h(0,b r).
\eeQ
Then
\beQ
\|\tilde{h}(0,r)\|_{X_{\infty,0}}=\sup_{r\geq 0}{\Blb\Big(1+\frac{r}{b}\Big)^{1+\ep}\Bl|\breve{h}(r)|+\Ba b\frac{\pt \breve{h}}{\pr}(r)\Ba\Br\Brb}<\delta.
\eeQ
Thus there exists a global classical solution $\tilde{h}(u,r)$ of (\ref{3}) in $I_0$ with initial data $\tilde{h}(0,r)$. By scaling group invariance of (\ref{3}) \cite{C1},
\beQ
h(u,r)=\tilde{h}\Big(\frac{u}{a},\frac{r}{a}\Big)
\eeQ
is also a global classical solution of (\ref{3}) in $I_0$ with initial data $h(0,r)$. Furthermore, $h$ satisfies
\begin{equation*}
|h(u,r)|\leq \frac{C}{\Big(1+\frac{u}{2}+r\Big)^{1+\ep}},\quad\Ba\frac{\pt h}{\pr}(u,r)\Ba\leq\frac{C}{\Big(1+\frac{u}{2}+r\Big)^{1+\ep}},
\end{equation*}
for some constant $C$.

Next, using the similar argument as the proof of Lemma \ref{lemA}, we can show
\beq\label{h-bh}
|h-\bh|\leq \frac{C_1 r}{\Big(1+\frac{u}{2}\Big)^{\ep-1} \Big(1+\frac{u}{2}+r \Big)^2},\label{b1}
\eeq
for some constant $C_1=C_1(C,\ep)$. This implies that
\beQ
\int_{r}^{\ift}{(h-\bh)^2\frac{dr'}{r'}}\leq\frac{C_1^2}{2\Big(1+\frac{u}{2}\Big)^{2\ep-2}\Big(1+\frac{u}{2}+r\Big)^2}.\label{b2}
\eeQ
Therefore
\beQ
g(u,r)=e^{2\beta}\geq\exp\Big[-\frac{2\pi C_1^2}{\Big(1+\frac{u}{2}\Big)^{2\ep-2}\Big(1+\frac{u}{2}+r\Big)^2}\Big].\label{b3}
\eeQ
Using the inequality $e^{-x}\geq 1-x$ for $x\geq 0$, we obtain
\beQ
0 \leq 1-g(u,r)\leq \frac{2\pi C_1^2}{\Big(1+\frac{u}{2}\Big)^{2\ep-2}\Big(1+\frac{u}{2}+r\Big)^2}.\label{b4}
\eeQ
Thus,
\beQ
\begin{aligned}
r(1-g)\leq & \frac{2\pi C_1^2 r}{\Big(1+\frac{u}{2}\Big)^{2\ep-2}  \Big(1+\frac{u}{2}+r\Big)^2}\\
\leq & \frac{2\pi C_1^2}{\Big(1+\frac{u}{2}\Big)^{2\ep-2} \Big(1+\frac{u}{2}+r\Big)}.
\end{aligned}
\eeQ
This implies
\beq
\lim\limits_{r\rw \ift}{r(1-g)}=0.\label{b6}
\eeq
Therefore
\beQ
\lim\limits _{r\rw\ift}{g}=1, \quad \lim\limits_{r\rw\ift}{g^{-1}}=1.
\eeQ
From (\ref{bondi}), (\ref{C-H}), we obtain
\beQ
m(u,r)=\frac{1}{g}\Big[\frac{r}{2}(1-\bg)-\frac{r}{2}(1-g)\Big]=\frac{1}{g}\Big[m_B(u)-\frac{r}{2}(1-g)\Big].
\eeQ
In terms of (\ref{b6}), we obtain
\beQ
M(u)=\lim_{r\rw\ift}{m(u,r)}= \lim_{r\rw\ift}{m_B(u,r)}=M_B (u).
\eeQ
Thus, for wave-like decaying solutions, the Bondi-Christodoulou mass is equal to the Bondi mass.

Now similar to the proof of Lemma \ref{lemA}, we can show
\beQ
|\bh|\leq\frac{1}{r}\ir{h dr'}\leq\frac{2C}{\ep \Big(1+\frac{u}{2}\Big)^{\ep}  \Big(1+\frac{u}{2}+r \Big)}.\label{b10}
\eeQ
Then
\beQ
|h-\bh|\leq |h|+|\bh|\leq \frac{C'}{\ep \Big(1+\frac{u}{2}\Big)^{\ep} \Big(1+\frac{u}{2}+r \Big)},\label{b11}
\eeQ
where $C'=\frac{2C}{\ep}+C$. From (\ref{m}), we obtain
\beQ
\begin{aligned}
0\leq m(u,r) \leq & \frac{2\pi {C'} ^2}{\ep^2 \Big(1+\frac{u}{2}\Big)^{2\ep}}\int_{0}^{r}{\frac{dr'}{\Big(1+\frac{u}{2}+r' \Big)^2}}\\
=&\frac{2\pi {C'} ^2 r}{\ep^2 \Big(1+\frac{u}{2}\Big)^{2\ep +1}\Big(1+\frac{u}{2}+r \Big)}.
\end{aligned}
\eeQ

Let $r\rw\ift$, we get
\beQ
0 \leq M_B(u)=M (u)=\lim_{r\rw\ift}{m(u,r)}\leq \frac{2\pi {C'} ^2}{\ep^2 \Big(1+\frac{u}{2} \Big)^{2\ep+1}}.
\eeQ
This implies that the final Bondi mass
\beQ
M_{B1} =\lim_{u\rw\ift}{M_B (u)}=0.
\eeQ
The uniqueness can be proved as same as Theorem \ref{thm1}.

Finally, we adopt the argument in \cite{LOY} to prove the completeness of causal geodesics of the global classical solution. The following lemmas are indeed Lemma 6.1-Lemma 6.11 in \cite{LOY}. We provide them here as there are slight differences on the proofs because of using different coordinate systems.

Let $I$ be an interval and $$\gamma : I\rw \mathscr{M} $$ be a future pointing causal geodesic on the above spacetime $\mathscr{M}$.
For any function $f$ on $\mathscr{M}$, its restriction on $\gamma$ and its derivatives are
\beQ
f(s)=f(\gamma(s)),\quad \dot{f}(s)=\frac{d}{ds}f(s),\quad \ddot{f}(s)=\frac{d^2}{ds^2}f(s).
\eeQ
Under the Bondi-Sachs coordinate system, we can write
\beQ
\gamma(s)=\Big(u(s),r(s),\theta(s),\psi(s)\Big),\quad \dot{\gamma}(s)=\Big(\dot{u}(s),\dot{r}(s),\dot{\theta}(s),\dot{\psi}(s)\Big).
\eeQ
Note that these are only defined away from the axis $\Gamma=\{r=0\}$. Same as in \cite{LOY}, we denote
\beQ
\begin{aligned}
& \mathbf{C}^2=-g_{\alpha\beta}|_{\gamma(s)}\dg^{\alpha}(s)\dg^{\beta}(s), \\
& \, \mathbf{J}^2=r^4   \big(g_{S^2,\theta\theta}\dot{\theta}^2+g_{S^2,\psi\psi}\dot{\psi}^2 \big).
\end{aligned}
\eeQ
As pointed out in \cite{LOY}, $\mathbf{C}^2 $, $\mathbf{J}^2$ are conserved and $ \mathbf{C}^2 >0$ when $\gamma(s)$ is a time-like geodesic. Let
\beq\label{g4}
U=\bar g\dot{u}+2\dot{r}
\eeq
Then
\beq
g\dot{u}U=\mathbf{C}^2+r^{-2}\mathbf{J}^2. \label{g5}
\eeq
For future pointing causal geodesics,
\beQ
\dot{u}\geq 0,\quad U \geq 0.
\eeQ

Note that the line $r=0$ is complete. This is because
\beQ
g, \,\bar g \geq k>0 \Lrw \int _0 ^u (g \bar g) ^\frac{1}{2}(u,0) du \geq k u  \lrw \infty
\eeQ
as $u \rw \infty$. Moreover, let
\beq\label{g6}
E(s)=g\bar g \dot{u}+g\dot{r}=\frac{1}{2}\big( g\bar g\dot{u}+g U \big).
\eeq
It is clearly that, away from the axis $\Gamma$, $E(s)$ is nonnegative. Thus Lemma 6.1, Lemma 6.2, Lemma 6.3 and Lemma 6.4 in \cite{LOY} hold true also in the current case. They are

\begin{lem}\label{G2}
Any future pointing causal geodesic $\gamma :[0,s_f)\rw\mathscr{M}$ can be continued past $s_f$ if there exists a compact subset $K\subseteq \mathscr{M}$ such that
\beQ
\{\gamma(s)\in\mathscr{M}: s\in [0,s_f)\}\subseteq K.
\eeQ
\end{lem}

\begin{lem}\label{G3}
If $\gamma(s): [0,s_f)\rw\mathscr{M}$ is incomplete, then either $\mathbf{C}\neq 0$ or $\mathbf{J}\neq 0$.
\end{lem}

\begin{lem}\label{G4}
If $\gamma: [0,s_f)\rw\mathscr{M}$ is incomplete, then set $\{s:r(s)=0\}$ is a discrete subset of $[0,s_f)$ (with a possible accumulation point at $s_f$).
\end{lem}

\begin{lem}\label{G5}
Let $\gamma: [0,s_f)\rw\mathscr{M}$ be a future geodesic with either $\mathbf{C}\neq 0$ or $\mathbf{J}\neq 0$, then $E(s)>0$ for $s\in [0,s_f)$.
\end{lem}

Now we derive the equations of each parameter of geodesic equations.

\begin{lem}\label{G6}
If geodesic $\gamma$ lies outside the axis $\Gamma$, then
\beq
\begin{aligned}
\ddot{u}(s)=& \Big[-g^{-1}g_u+\frac{1}{2}g^{-1}(g\bar g)_r\Big]\dot{u}^2-g^{-1}r^{-3}\mathbf{J}^2,\\
\ddot{r}(s)=& g^{-1} \Big[-\frac{1}{2}(g\bar g)_u+\bar g g_u-\frac{1}{2}\bar g (g\bar g)_r\Big]\dot{u}^2\\
            & - g^{-1} \Big[(g\bar g)_r\dot{u}\dot{r} +g_r\dot{r}^2 -\bar g r^{-3}\mathbf{J}^2\Big],\\
  \dot U(s)=&-\frac{1}{2}g^{-1}(g\bar g)_r\dot uU-g^{-1}g_rU\dot r+g^{-1}\bar g r^{-3}\mathbf{J}^2,\\
 \dot{E}(s)=&\frac{1}{2}(g\bar g)_u\dot{u}^2+g_u\dot{u}\dot{r}=\frac{1}{2}\big( g_u\dot{u}U+ g\bar g_u\dot{u}^2 \big),
\end{aligned}
\label{g7}
\eeq
where $f_u$, $f_r$ are the partial derivatives of $f$ with respect to $u$ and $r$.
\end{lem}
\pf Substitute the Christoffel symbols in Appendix into the geodesic equations
\beQ
\ddot{\gamma}^{\lambda}=-\Gamma_{\alpha\beta}^{\lambda}\dot{\gamma}^{\alpha}\dot{\gamma}^{\beta},
\eeQ
we can derive the first two equations. They imply
\beQ
\begin{aligned}
\dot U(s)=&\bar g_u\dot u^2+\bar g_r\dot u\dot r+\bar g\ddot u+2\ddot r\\
=&-\frac{1}{2}g^{-1}(g\bar g)_r\dot uU-g^{-1}g_rU\dot r+g^{-1}\bar g r^{-3}\mathbf{J}^2,\\
\dot{E}(s)=&\dot{g}\bar g\dot{u}+g\dot{\bar g}\dot{u}+g\bar g\ddot{u}+\dot{g}\dot{r}+g\ddot{r}\\
=&(g_u\dot{u}+g_r\dot{r})\bar g\dot{u}+g(\dot{\bar g}_u\dot{u}+\dot{\bar g}_r\dot{r})\dot{u}+g\bar g\ddot{u}\\
&+(g_u\dot{u}+g_r\dot{r})\dot{r}+g\ddot{r}\\
=&\frac{1}{2}(g\bar g)_u\dot{u}^2+g_u\dot{u}\dot{r}\\
=&\frac{1}{2}\big(g_u\dot{u}U+g\bar g_u\dot{u}^2 \big).
\end{aligned}
\eeQ
\qed

\begin{lem}\label{G7}
If $\gamma(s):[0,s_f)\rw \mathscr{M}$ is incomplete with $\mathbf{C}\neq 0$ or $\mathbf{J}\neq 0$, then there exists some constant $C>0$ such that,
for any $s \in [0,s_f)$,
\beq\label{g8}
E(s)\geq\frac{C}{s_f-s}.
\eeq
\end{lem}
\pf Since
\beQ
1 \geq g,\,\, \bar g \geq k>0,
\eeQ
we have
\beQ
\dot{u}+U\leq \frac{2}{k^2}E(s).
\eeQ
In the following we denote $C_1$, $C_2$ and $C_3$ are certain positive constants. As the solution and its derivative satisfy
\beQ
|h_u|,  \,\, |\bar h_u|\leq C_1.
\eeQ
Thus, using (\ref{h-bh}), we obtain
\beQ
|g_u|\leq 4\pi|g|\int_{r}^{\infty}{2|h-\bar h| |h_u-\bar h_u|\frac{dr'}{r'}}\leq C_2,
\eeQ
\beQ
|\bar g_u|\leq \frac{1}{r}\int_{0}^{r}{|g_u| dr'}\leq C_2.
\eeQ
Therefore
\beQ
\dot{E}(s)\leq 2C_3 \big(\dot{u}^2+U^2 \big)\leq \frac{8C_3}{k^4}E^2(s).
\eeQ
Then the lemma follows by applying the same argument as proving Lemma 6.6 in \cite{LOY}. \qed

\begin{lem}\label{G8}
If $\gamma(s):[0,s_f)\rw \mathscr{M}$ is incomplete, then for any $r_0>0$ and any $s_0\in[0,s_f)$, there exists $s\in [s_0,s_f)$ such that $r(s)<r_0$.
\end{lem}
\pf We prove it by contradiction. If there exist some constant $r_0>0$ and $s_0\in[0,s_f)$ such that $r(s)\geq r_0$ for all $s\in[s_0,s_f)$, then
Lemma \ref{G6} gives
\beQ
\begin{aligned}
\frac{d}{ds}(g U)=&(g\bar g)_u\dot u^2+(g\bar g)_r\dot u\dot r+(g\bar g)\ddot u+2g_u\dot u\dot r+2g_r\dot r^2+2g\ddot r\\
=&\Big[(g\bar g)_u-(g\bar g)g^{-1}g_u+\frac{1}{2}(g\bar g)g^{-1}(g\bar g)_r -(g\bar g)_u+2 \bar g g_u \\
 &-\bar g(g\bar g)_r \Big]\dot{u}^2 +\Big[2g_u-(g\bar g)_r \Big]\dot u\dot r+\bar gr^{-3}\mathbf{J}^2\\
=&\Big[g_u-\frac{1}{2}(g\bar g)_r \Big]\Big( \mathbf{C}^2+r^{-2}\mathbf{J}^2 \Big)+\bar g r^{-3}\mathbf{J}^2.
\end{aligned}
\eeQ
As $\mathbf{C}^2$ and $\mathbf{J}^2$ are conserved, $g$, $\bar g$ and their derivatives are uniformly bounded, there exists constant $C_4>0$ such that
\beQ
C_4 \geq U \geq 0.
\eeQ
This gives
\beQ
C_4-k\dot u-2\dot r \geq 0.
\eeQ
Integrating it from $s_0$ to $s$, we obtain
\beQ
C_4 (s-s_0)-ku(s)+ku(s_0)-2r(s)+2r(s_0) \geq 0.
\eeQ
Thus,
\beQ
u(s)\leq \frac{1}{k}\Big[C_4 (s_f-s_0)+ku(s_0)-2r_0+2r(s_0)\Big].
\eeQ
As $\dot u\geq 0$, we conclude that $u$ is uniformly bounded. Similarly, $r$ is also uniformly bounded. Thus $\gamma(s)$ lies in a compact set in $\mathscr{M}$. This contradicts Lemma \ref{G2} and the proof of lemma is complete. \qed

The following lemma and its proof are the same as Lemma 6.8 in \cite{LOY}.

\begin{lem}\label{G9}
Assume $\gamma(s):[0,s_f)\rw\mathscr{M}$ is incomplete. Then $\mathbf{J}\neq 0$.
\end{lem}

\begin{lem}\label{G10}
There exists $r_0>0$ such that if $\mathbf{J} \neq 0$ and at some time $s_0$,
\beQ
\dot u(s_0),\,\, U(s_0) \leq \frac{2}{k r(s_0)}\mathbf{J},\quad \dot r(s_0)\leq 0,\quad r(s_0)<r_0,
\eeQ
where $k$ is defined by (\ref{kk}), then
\beQ
\ddot u(s_0)<0, \quad \dot U(s_0)>0.
\eeQ
Moreover, if $\dot r(s_0)=0$, then $$\ddot r(s_0)>0.$$
\end{lem}
\pf It is straightforward that
\beQ
|g_r|,\,\,|g_u|,\,\,|\bar g_r|,\,\,|\bar g_u|\leq C_5
\eeQ
for some constant $C_5 >0$. Since both $\mathbf{C}^2$ and $\mathbf{J}^2$ are conserved, there exists $\tilde{r}$ such that for any $0<r\leq \tilde{r}$
\beQ
\mathbf{C}^2+\frac{1}{r^2}\mathbf{J}^2\leq \frac{2}{r^2}\mathbf{J}^2.
\eeQ
Take
\beQ
r_0=\min{\Big\{\frac{k^4}{16 C_5},\,\,\tilde{r}\Big\}}.
\eeQ
Then (\ref{g7}) gives
\beQ
\begin{aligned}
g\ddot u(s_0)&=\Big[-g_u+\frac{1}{2}(g\bar g)_r \Big]\dot u^2-r^{-3}\mathbf{J}^2\\
&\leq \Big(|g_u|+\frac{1}{2}|g||\bar g_r|+\frac{1}{2}|g_r||\bar g| \Big)\frac{4}{k^2 r^2(s_0)}\mathbf{J}^2-\frac{1}{r^3(s_0)}\mathbf{J}^2\\
&\leq \frac{8 C_5}{k^2 r^2(s_0)}\mathbf{J}^2- \frac{16 C_5}{k^4 r^2(s_0)} \mathbf{J}^2\\
&\leq -\frac{8 C_5}{k^4 r^2(s_0)}\mathbf{J}^2 <0.
\end{aligned}
\eeQ
Therefore
\beQ
\ddot u(s_0)<0.
\eeQ
The remaining conclusions can be proved in the same way by using (\ref{g7}). \qed

\begin{lem}\label{G11}
Assume $\gamma(s):[0,s_f)\rw \mathscr{M}$ is incomplete and $\mathbf{J}\neq 0$. Then there exists $r_0>0$ such that for every $s_0\in[0,s_f)$ the geodesic $\gamma(s)$ exits the cylinder with radius $r_0$ at some time to the future of $s_0$, that is, there exists $s_1\in(s_0,s_f)$ such that $r(s_1)>r_0$.
\end{lem}
\pf Let $r_0$ be given in Lemma \ref{G10}. Suppose the geodesic $\gamma(s)$ lies in the cylinder with radius $r_0$ for $s\in[s_0,s_f)$.\\\\
\textsl{Step 1.} We claim that, for any $s\in[s_0,s_f)$,
\beq
\dot{r}(s)\leq 0.   \label{g9}
\eeq
If there exists $s'\in[s_0,s_f)$ such that
\beQ
\dot r(s')>0,
\eeQ
then Lemma \ref{G8} implies that there exists $s''>s'$ such that
\beQ
\dot r(s'')<0.
\eeQ
Take
\beQ
s^{\ast}=\sup{\Big\{s:s'\leq s\leq s'',\,\,\dot r(s)\geq 0\Big\}}.
\eeQ
Then
\beQ
\dot r(s^{\ast})=0
\eeQ
and, for $s^{\ast}<s\leq s''$,
\beQ
\dot r(s)<0.
\eeQ
Thus (\ref{g4}) implies
\beQ
k\dot u(s^{\ast})\leq U(s^{\ast})=\bar g\dot u(s^{\ast})\leq \dot u(s^{\ast})
\eeQ
and (\ref{g5}) implies
\beQ
\frac{k^2}{2}\leq g\dot u^2(s^{\ast})\leq C^2+ \frac{1}{r^2 (s^{\ast})} \mathbf{J}^2 \leq \frac{2}{r^2(s^{\ast})}\mathbf{J}^2.
\eeQ
Therefore, we obtain
\beQ
U(s^{\ast})\leq \dot u(s^{\ast})\leq \frac{2}{k r(s^{\ast})}\mathbf{J}.
\eeQ
By Lemma \ref{G10}, we have
\beQ
\ddot r (s^{\ast})>0 \Lrw \dot r (s) >0, \quad s>s^\ast.
\eeQ
This gives contradiction. Hence (\ref{g9}) holds.\\\\
\textsl{Step 2.} We show that there exists $t_0\in(s_0,s_f)$ such that
\beq\label{g10}
\frac{k}{2}\dot u(t_0)\leq U(t_0).
\eeq
If not, for all $s\in(s_0,s_f)$
\beQ
\frac{k}{2}\dot u>U,
\eeQ
then
\beQ
\dot r(s)=\frac{1}{2}(U-\bar g\dot u)\leq -\frac{k}{4}\dot u(s).
\eeQ
Integrating it from $s_0$ to $s$, we obtain that $u$ is uniformly bounded, which violates Lemma \ref{G2}.
Hence (\ref{g10}) holds.\\\\
\textsl{Step 3.} Let $t_0$ be given in (\ref{g10}), we claim that, for $s\in[t_0,s_f)$,
\beq\label{g11}
\frac{k}{2}\dot u(s)\leq U(s).
\eeq
Define
\beQ
s^{\ast}=\sup{\Big\{s:\frac{k}{2}\dot u(s')\leq U(s'),\,\,\forall s'\in[t_0,s]\Big\}}.
\eeQ
If $s^{\ast}=s_f$, then (\ref{g11}) holds. If not, by continuity,
\beQ
\frac{k}{2}\dot u(s^{\ast})\leq U(s^{\ast}).
\eeQ
Since
\beQ
\dot r(s)\leq 0,
\eeQ
we have
\beQ
U(s^{\ast})\leq \dot u(s^{\ast}).
\eeQ
Similar to \textsl{Step 1}, we can show that $\dot u(s^{\ast})$ and $U(s^{\ast})$ satisfy the conditions in Lemma \ref{G10}. Then
\beQ
\ddot u(s^{\ast})<0, \quad \dot U(s^{\ast})>0.
\eeQ
Thus, there exists $t_1>s^{\ast}$ such that, for all $s\in[s^{\ast},t_1]$,
\beQ
\frac{k}{2}\dot u(s)\leq \frac{k}{2}\dot u(s^{\ast})\leq U(s^{\ast})\leq U(s).
\eeQ
This contradicts the definition of $s^{\ast}$. Hence (\ref{g11}) holds.\\\\
\textsl{Step 4.} The above arguments and Lemma \ref{G10} imply that, for all $s\in[t_0,s_f)$.
\beQ
\ddot u(s)<0,\quad U(s)\leq \dot u(s).
\eeQ
Therefore $\dot u$ and $U(s)$ are uniformly bounded. This contradicts Lemma \ref{G2}. Thus the proof of lemma is complete.\qed

\begin{lem}\label{G12}
Assume $\gamma(s):[0,s_f)\rw \mathscr{M}$ is incomplete and $r(s)>0$ for all $s\in[0,s_f)$. Suppose there exists a sequence $\{s_n\}$ with $s_n\rw s_f$ such that $\dot r(s_n)=0$. Then
\beQ
\lim\limits_{n\rw \infty}{(\dot uU )(s_n)}=\infty.
\eeQ
\end{lem}
\pf By (\ref{g4}), we have
\beQ
k\dot u(s_n)\leq U(s_n)\leq \dot u(s_n).
\eeQ
Then (\ref{g6}) gives
\beQ
E(s_n)\lesssim \sqrt{(\dot uU)}(s_n).
\eeQ
Therefore the conclusion follows from Lemma \ref{G7} and the proof of lemma is complete.\qed

Now we prove future geodesic completeness. If $\gamma(s)$ is incomplete, then Lemma \ref{G9} indicates $\mathbf{J}\neq 0$. Take $r_0$ sufficiently small such that Lemma \ref{G11} holds. Lemma \ref{G8} and Lemma \ref{G11} imply that $\gamma(s)$ intersects the cylinder $r=r_0$ infinitely many times. Thus we can find a sequence $s_n\rw s_f$ satisfying the conditions in Lemma \ref{G12}. As (\ref{g5}) shows that $\dot u(s_n)$ and $U(s_n)$ are uniformly bounded. This contradicts Lemma \ref{G12}. So the proof of theorem is complete.\qed

\begin{rmk}
The past causal geodesics are also complete by reversing the time-orientation.
\end{rmk}

\mysection{Generalized solutions for large data}
\ls

In this section, we generalize Christodoulou's generalized solutions of (\ref{3}) to the wave-like decaying condition (\ref{wave-like}). As most proofs in \cite{C2, C3} can go through without any change, we only provide the proof of Lemma \ref{lemD}.

The $\al$-regularized equation is given as follows
\beq
D_{\al}\ha=\frac{1}{2(r+\al)}(\ga-\bga)(\ha-\bha)\label{502}
\eeq
where
\beQ
\begin{aligned}
\bha=&\frac{1}{r+\al}\ir{\ha dr'}, \\
\ga=&\exp\Big[-4\pi \int_{r}^{\ift}{(\ha-\bha)^2\frac{dr'}{r'}}\Big], \\
\bga=&\frac{1}{r+\al}\ir{\ga dr'}.
\end{aligned}
\eeQ
Clearly,
\beQ
\bga(u,0)=0.
\eeQ

Denote the differential operator
\beQ
D_{\al}=\frac{\pt}{\pu}-\frac{1}{2}\bga\frac{\pt}{\pr}.\label{d6}
\eeQ
The integral curve of $D_{\al}$, denoted by $r=\chi_{\al}$, satisfies the ODE
\beQ
\frac{dr}{du}=-\frac{1}{2}\bga.\label{d7}
\eeQ
Define the $\alpha$-local mass function
\beQ
m_{\al}=\frac{r+\al}{2}\Bl1-\frac{\bga}{\ga}\Br.
\eeQ

From \cite{C2}, we know that $m_{\al}$ is a monotonically nondecreasing function with respect to $r$ and we can derive
\beQ
m_{\al}=\frac{\al}{2}+2\pi\ir{\frac{\bga}{\ga}(\ha-\bha)^2dr'}.
\eeQ
Then the $\alpha$-total mass $M_{\al}(u)$ can be defined \cite{C2}
\beQ
M_{\al}(u)=\lim\limits_{r\rw\ift}{m_{\al}(u,r)}.
\eeQ

It is straightforward that
\beQ
D_{\al}\bha=\frac{\xi_{\al}}{2(r+\al)},\quad D_{\al}m_{\al}=-\frac{\pi}{\ga}\xi_{\al}^2,
\eeQ
where
\beQ
\xi_{\al}=\ir{\bga(\ha-\bha)\frac{dr'}{r'+\al}}.
\eeQ
Then the following identity holds
\beQ
\begin{aligned}
\int_{\delta}^{r_1} & {\frac{\ga}{\bga}(u_1,r)dr}
+2\pi\iint\limits_{I_{\delta,\al}(u_1,r_1)}{\frac{\ga}{\bga^2}\frac{\xi_{\al}^2}{r+\al}drdu}\\
&+\frac{1}{2}\int_{0}^{u_1}{\ga(u,\delta)du}=\int_{\delta}^{r_{0,\al}}{\frac{\ga}{\bga}(0,r)dr},
\end{aligned}
\eeQ
where $r_{0,\al}=\chi_{\al,u_1}(0;r_1)$,
$$I_{\delta,\al}(u_1,r_1)=\Big\{(u,r)|0<u<u_1,\delta<r<\chi_{\al,u_1}(u;r_1)\Big\}.$$

\begin{lem}\label{lemD}
For any initial data $\breve{h}(r) \in C^1[0,\ift)$ which satisfies (\ref{wave-like}), then, for any $\al>0$, there exists unique global classical solution
\beQ
\ha(u,r)\in C^1\Big([0,\ift)\times[0,\ift)\Big)
\eeQ
of (\ref{502}) which satisfies the initial condition $h(0,r)=\breve{h}(r)$ and the decay property (\ref{wave-like}) at each $u\geq 0$.
\end{lem}
\pf For any $\al>0$, the existence of unique global classical solution can be proved by the same argument as in \cite{C2}. In the following we prove that the solution preserves the wave-like decay at null infinity.

Let $\chi_{\alpha}(u;r_1)$ be the characteristic through $(u_1,r_1)$ and
\beQ
r_{0,\alpha}=\chi_{\alpha}(0;r_1).
\eeQ
Integrating (\ref{502}) along $\chi_{\alpha}(u;r_1)$, we obtain
\beq
\begin{aligned}
h_{\alpha}(u_1,r_1)=h(0,r_{0,\alpha})+\int_{0}^{u_1} \Big[\frac{g_{\alpha}-\bar{g}_{\alpha}}{2(r+\alpha)}(h_{\alpha} -\bar{h}_{\alpha})\Big]_{\chi_{\alpha}}du.
\end{aligned}
\label{503}
\eeq

Let $\ma=M_{\al}(0)$, and
\beQ
x(u)=\sup\limits_{r\geq 4\ma}{\Big[\Big(\frac{r}{4\ma}\Big)^{1+\ep}|h_{\alpha}(u,r)|\Big]}.
\eeQ
As pointed out in \cite{C2}, for any differentiable function $f$
\beQ
\begin{matrix}
\left\{
\begin{aligned}
& f, \,\, r\dfrac{\partial f}{\partial r} \in L^2(0,\ift)\\
& \lim\limits_{r\rw\ift}{rf^2(r)}=0
\end{aligned}
\right.
\end{matrix}
\Lrw r_1f^2(r_1)\leq \int_{r_1}^{\ift}{r^2\Big(\frac{\pt f}{\pt r}\Big)^2dr}.
\eeQ
Taking $r_1=4\ma$ and $f=\bar{h}_{\alpha}$ in the inequality, we obtain
\beQ
\begin{aligned}
4\ma\bar{h}^2 _{\al}(4\ma)\leq & \int_{4\ma}^{\ift}{(r+\alpha)^2\Big(\frac{\pt \bar{h}_{\alpha}}{\pr}\Big)^2dr}\\
                          = & \int_{4\ma}^{\ift} (h_\alpha -\bar h _\alpha )^2 dr\leq \frac{\ma}{\pi}.
\end{aligned}
\eeQ
This gives
\beQ
\bar{h}^2 _{\al}(4\ma)\leq \frac{1}{2 \pi ^\frac{1}{2}}.
\eeQ
Thus, for $r\geq 4\ma$,
\beq
\begin{aligned}
|\bar{h}_{\alpha}(r)|&=\frac{1}{r+\alpha}\Ba(4\ma+\alpha)\bar{h}_{\alpha}(4\ma)+\int_{4\ma}^{r}
{h_{\alpha}(r')dr'}\Ba\\
&\leq \frac{1}{r+\alpha}\Big[\frac{4\ma+\alpha}{2\pi^{\frac{1}{2}}}+\int_{4\ma}^{r}{x\Big(\frac{4\ma}{r'}\Big)^{1+\ep}dr'}\Big]\\
&\leq \Big[\Big(1+\frac{\alpha}{4\ma}\Big)\frac{1}{2\pi^{\frac{1}{2}}}+\frac{x}{\ep}\Big]\frac{4\ma}{r}.
\end{aligned}
\label{504}
\eeq
On the other hand,
\beq
\frac{g_{\al}-\bga}{2(r+\al)}=\frac{m_{\al}}{(r+\al)^2}g_{\alpha}\leq\frac{\ma}{r^2}.\label{505}
\eeq
Thus $(\ref{503})$, $(\ref{504})$ and $(\ref{505})$ imply that
\beQ
\begin{aligned}
x(u) \leq & \sup\limits_{r\geq 4\ma}{\Big[\Big(\frac{r}{4\ma}\Big)^{1+\ep}|h_0(r)|\Big]}\\
&+\frac{1}{8\pi^{\frac{1}{2}}}\Big(1+\frac{\alpha}{4\ma}\Big)\frac{u}{4\ma}\\
&+\frac{\ep+1}{16\ep\ma}\int_{0}^{u}{x(u')du'}\\
\leq & e^{\tau} \sup\limits_{r\geq 4\ma}{\Big[\Big(\frac{r}{4\ma}\Big)^{1+\ep}|h_0(r)|\Big]}\\
& +\frac{e^{\tau}}{8\pi^{\frac{1}{2}}}\Big(1+\frac{\alpha}{4\ma}\Big)\frac{u}{4\ma}\Big],
\end{aligned}
\eeQ
where
\beQ
\tau=\frac{(1+\ep)u}{16\ep\ma}.
\eeQ
This implies the wave-like decaying for $h_\alpha$.

Next we show the wave-like decaying for $\frac{\pt h_{\al}}{\pr}$. This can be done by using the same argument to the following identity
(c.f. $(4.8)$ in \cite{C2}).
\beQ
\begin{aligned}
& \frac{\pt h_{\al}}{\pr}(u_1,r_1)=\frac{\pt h}{\pr}(0,r_{0,\alpha})+\int_{0}^{u_1}{\Big[\frac{g_{\al}-\bga}{r+\al}
  \frac{\pt h_{\al}}{\pr}\Big]_{\chi_{\al}}du}\\
& +\int_{0}^{u_1}{\Big\{\frac{1}{2(r+\al)^2}\Big[-3(\ga-\bga)+4\pi\ga(h_{\al}-\bha)^2 \Big](h_{\al}-\bha)\Big\}_{\chi_{\alpha}}du}.
\end{aligned}
\eeQ
Therefore, the proof of the lemma is complete. \qed

\begin{lem}\label{lem9}
For any $u_0, \,\delta>0$, three families of functions
\beQ
\Blb\ha\Ba\al\in (0,\frac{\delta}{2}]\Brb, \quad \Blb\bha\Ba\al\in (0,\frac{\delta}{2}]\Brb, \quad \Blb\dfrac{\pt\ha}{\pr}\Ba\al\in (0,\frac{\delta}{2}]\Brb
\eeQ
are equicontinuous in $[0,u_0]\times[\delta,\ift)$.
\end{lem}

\begin{lem}\label{lem5-3}
There exists a subsequence $\{\hai\}$ of the sequence $\{h_{\alpha}\}$ such that $\{\hai\}$ converges to a differentiable continuous function $h\in C^1(I)$ on each compact subset of $I=[0, \infty)\times (0, \infty)$ uniformly while the sequence $\Blb\dfrac{\pt \hai}{\pr}\Brb$ converges to $\dfrac{\pt h}{\pr}$ uniformly. Moreover, $h$ satisfies (\ref{3}) in $I$ and, for arbitrary $r_1>0$,
\beQ
h\in L^2(0,\ift), \quad \frac{g}{\bg}\in L^1(0,r_1).
\eeQ
\end{lem}

\begin{lem}\label{lem5-4}
At each $u$, the function
\beQ
\xi(r) =\lim\limits_{\delta\rw 0}{\int_{\delta} ^{r}{\frac{\bg(h-\bh)}{r'}}}dr'
\eeQ
exists. And, for arbitrary $r_0 >0$, the measurable function
\beQ
\dfrac{g^{\frac{1}{2}}\xi}{\bg r^{\frac{1}{2}}} \in L^2 \Big([0, \infty)\times (0, r_0]\Big).
\eeQ
\end{lem}

\begin{lem}\label{l1}
At almost all $u$, $\frac{\xi}{\bg^{\frac{1}{2}}}$ is a continuous function of $r$ and uniformly bounded such that
\beQ
\lim _{r \rw 0} \frac{\xi}{\bg^{\frac{1}{2}}} (u, r) = 0.
\eeQ
Also, for arbitrary $u_0>0$,
\beQ
\sup_{r\geq 0}{\Ba\frac{\xi}{\bg^{\frac{1}{2}}}(u,r)\Ba} \in L^2(0,u_0), \quad \lim _{r\rw 0}\int_{0}^{u_0}{\frac{\xi^2}{\bg}(u,r) du} =0.
\eeQ
\end{lem}

\begin{lem}\label{l2}
At almost all $u$,
\beQ
\lim _{\delta \rw 0} \Big(\delta DA(\delta) \Big)= 0.
\eeQ
\end{lem}

\begin{lem}\label{lem5-7}
$\bar{h}$ and $m$ are weakly differentiable in $I$ and
\beQ
D\bh=\frac{\xi}{2r}, \quad Dm=-\frac{\pi}{g}\xi^2.
\eeQ
\end{lem}

\begin{thm}
Given initial data $\breve{h}(r) \in C^1[0,\ift)$ which satisfies (\ref{wave-like}), there exists at least one global generalized solution which has the same data as a classical solution coincides with it in the domain of existence of the latter.
\end{thm}
\pf From the above lemmas, we know that there exists a solution $h\in C^1(I)$ which satisfies $(1)$-$(4)$ in Definition \ref{gensolu}.
By (5.41) in \cite{C2} we obtain
\beQ
D\Bl\int_{\delta}^{r}{\frac{g}{\bg}dr}\Br=-2\pi \int_{\delta}^{r}{\frac{g\xi^2}{r\bg^2}dr}-\frac{1}{2}g(\delta).
\eeQ
By using the dominated convergence theorem and letting $\delta \rw 0$, we know that $\ir{\frac{g}{\bg}dr}$ is weakly differentiable in $I$ and satisfies that
\beq
D\Bl\ir{\frac{g}{\bg}dr}\Br=-2\pi\ir{\frac{g\xi^2}{r\bg^2}dr}-\frac{1}{2}g(0).\label{509}
\eeq
Thus, integrating (\ref{509}) along the characteristic $\chi_{u_1}(u;r_1)$, we have
$$
\int_{0}^{r_1}{\frac{g}{\bg}(u_1,r)dr}+2\pi \iint\limits_{I(u_1,r_1)}{\frac{g\xi^2}{\bg^2r}drdu}+\frac{1}{2}\int_{0}^{u_1}{g(u,0)du}=-\int_{0}^{r_0}{\frac{g}{\bg}(0,r)dr}.
$$
This is indeed (5) in Definition \ref{gensolu}, so that global generalized solutions exist. The proof of uniqueness is the same as that in \cite{C3}. Therefore the proof of the theorem is complete. \qed

\mysection{Appendix: Spherically symmetric Bondi-Sachs metrics}\ls

The nontrivial metric components
\beQ
\begin{aligned}
& g_{uu}=-e^{2\beta}\frac{V}{r},\quad \,\,g_{ur}=-e^{2\beta}, \quad \,\,\,\,\,g_{\theta\theta}=r^2,\quad \,\,\,\, g_{\psi\psi}=r^2\sin^2\theta,\\
& g^{ur}=-e^{-2\beta},\quad \,\,\,\,g^{rr}=e^{-2\beta}\frac{V}{r},\quad g^{\theta\theta}=r^{-2},\quad \,g^{\psi\psi}=r^{-2}\sin^{-2}\theta.
\end{aligned}
\eeQ
The nontrivial Christoffel symbols
\begin{equation*}
\begin{aligned}
\Gamma_{uu}^u &=2\frac{\partial\beta}{\partial u}-\frac{V}{r}\frac{\partial\beta}{\partial r}-\frac{1}{2r}\frac{\partial V}{\partial r}+\frac{V}{2r^2},\\
\Gamma_{\theta\theta}^u &=re^{-2\beta},\quad \Gamma_{\psi\psi}^u =r\sin^2\theta e^{-2\beta},\\
\Gamma_{uu}^r &=\frac{V^2}{r^2}\frac{\partial\beta}{\partial r}+\frac{V}{2r^2}\frac{\partial V}{\partial r}-\frac{V^2}{2r^3}-\frac{V}{r}\frac{\partial \beta}{\partial u}+\frac{1}{2r}\frac{\partial V}{\partial u},\\
\Gamma_{ru}^r &=\frac{V}{r}\frac{\partial \beta}{\partial r}+\frac{1}{2r}\frac{\partial V}{\partial r}-\frac{V}{2r^2},\\
\Gamma_{rr}^r &=2\frac{\partial \beta}{\partial r},\quad \Gamma_{\psi\psi}^r=-Ve^{-2\beta}\sin^2\theta,\quad \Gamma_{\theta\theta}^r =-Ve^{-2\beta}\\
\Gamma_{\theta r}^{\theta} &=\frac{1}{r},\quad \Gamma_{\psi\psi}^{\theta}=-\sin\theta\cos\theta,\quad
\Gamma_{\psi r}^{\psi}=\frac{1}{r},\quad \Gamma_{\theta\psi}^{\psi}=\cot\theta.
\end{aligned}
\end{equation*}

The nontrivial components of the Riemann curvature tensors
\begin{equation*}
\begin{aligned}
R_{\theta r\theta}^u &=-2re^{-2\beta}\frac{\partial \beta}{\partial r},\\
R_{\theta r \theta}^r &=e^{-2\beta}\Big(\frac{\partial \beta}{\partial r}-\frac{1}{2}\frac{\partial V}{\partial r}+\frac{V}{2r}\Big),\\
R_{uru}^u &=2\frac{\partial^2\beta}{\partial r\partial u}-\frac{V}{r}\frac{\partial^2\beta}{\partial r^2}-\frac{V}{r^2}-\frac{V}{2r^3}-\frac{1}{r}\frac{\partial V}{\partial r}+\frac{1}{2rV}(\frac{\partial V}{\partial r})^2,\\
R_{uru}^r &=-\frac{2V}{r}\frac{\partial^2\beta}{\partial r\partial u}+\frac{V^2}{r^2}\frac{\partial^2\beta}{\partial r^2}+\frac{V}{r^2}\frac{\partial \beta}{\partial r}\frac{\partial V}{\partial r}-\frac{V^2}{r^3}\frac{\partial \beta}{\partial r}-\frac{V}{r^3}\frac{\partial V}{\partial r}+\frac{V^2}{r^4},\\
R_{u\theta u}^{\theta}&=R_{u\psi u}^{\psi}=\frac{V^2}{r^3}\frac{\partial \beta}{\partial r}+\frac{V}{2r^3}\frac{\partial V}{\partial r}-\frac{V}{r^2}\frac{\partial \beta}{\partial u}-\frac{1}{2r^2}\frac{\partial V}{\partial u}-\frac{V^2}{2r^4},\\
R_{r\theta r}^{\theta}&=R_{r\psi r}^{\psi}=\frac{2}{r}\frac{\partial \beta}{\partial r},\\
R_{u\theta r}^{\theta}&=R_{u\psi r}^{\psi}=\frac{V}{r^2}\frac{\partial \beta}{\partial r}+\frac{1}{2r^2}\frac{\partial V}{\partial r}-\frac{V}{2r^3},\\
R_{\psi\theta\psi}^{\theta}&=\sin^2\theta \Big(1-\frac{V}{r}e^{-2\beta}\Big).\\
\end{aligned}\end{equation*}

The nontrivial components of the Ricci curvature tensors
\begin{equation*}
\begin{aligned}
R_{uu}&=-\frac{2V}{r}\frac{\partial^2\beta}{\partial u\partial r}+\frac{V^2}{r^2}\frac{\partial^2\beta}{\partial r^2}+\frac{V}{r^2}\frac{\partial \beta}{\partial r}\frac{\partial V}{\partial r}+\frac{V^2}{r^3}\frac{\partial \beta}{\partial r}-\frac{V}{r}\frac{\partial \beta}{\partial u}+\frac{1}{2r}\frac{\partial V}{\partial u},\\
R_{ur}&=-2\frac{\partial^2\beta}{\partial u\partial r}+\frac{V}{r}\frac{\partial^2\beta}{\partial r^2}+\frac{1}{r}\frac{\partial \beta}{\partial r}\frac{\partial V}{\partial r}+\frac{V}{r^2}\frac{\partial \beta}{\partial r},\\
R_{rr}&=\frac{4}{r}\frac{\partial \beta}{\partial r},\\
R_{\theta\theta}&=\sin^{-2}\theta R_{\psi\psi}=1-e^{-2\beta}\frac{\partial V}{\partial r}.
\end{aligned}\end{equation*}

{\footnotesize {\it Acknowledgement. The authors would like to thank Qing Han, Shiwu Yang and Pin Yu, as well as referees for some valuable suggestions. This work is supported by Chinese NSF grants 11571345, 11731001, the special foundations for Guangxi Ba Gui Scholars and Junwu Scholars of Guangxi University.}}

\end{document}